\documentclass{aa} 
\usepackage{float}
\usepackage{tikz}
\usepackage{comment}
\usetikzlibrary{shapes,arrows}
\usepackage{soul}
\usepackage{graphicx}

\bibpunct{(}{)}{;}{a}{}{,} % to follow the A&A style
\usepackage[varg]{txfonts}
\usepackage[bookmarks=true,
  colorlinks,
  linkcolor=blue,
  urlcolor=blue,
  citecolor=blue,
  plainpages=false,
  pdfpagelabels,
  final,
  breaklinks=true]{hyperref}
\usepackage{mathtools}
\usepackage{cancel}

\newcommand{\ct}[1]{{\color{blue}Triantafyllou et al. in prep}}

\begin{document}

\title{TILING I: Field-level Bayesian reconstruction of cosmological initial conditions during the epoch of reionization}

\author{Nikolaos Triantafyllou
      \inst{1}\thanks{nikolaos.triantafyllou@sns.it}\orcid{0009-0003-8609-4529}
      \and
      Andrei Mesinger\inst{2, 1}\orcid{0000-0003-3374-1772}
     \and 
    Steven Murray\inst{3, 1} \orcid{0000-0003-3059-3823}
    \and
    Samuel Gagnon-Hartman\inst{1} \orcid{0009-0006-7603-644X}
      }
      
\authorrunning{N. Triantafyllou et al.}
\titlerunning{TILING I: Reconstruction of ICs during the EoR}

\institute{Scuola Normale Superiore, Piazza dei Cavalieri 7, 56125 Pisa, PI, Italy           \and
            Department of Physics and Astronomy {\it ``Ettore Majorana''}, University of Catania, Via Santa Sofia 64, 95123  Catania, Italy
             \and
             Physics Department, Stellenbosch University, 42 Merriman Ave, Stellenbosch, South Africa, 7600
          }

\date{}

% \abstract{}{}{}{}{}
% 5 {} token are mandatory
\abstract
{
Reconstructing the initial conditions (ICs) of the matter field corresponding to a given observed volume provides a complete picture of the temporal evolution of that piece of our Universe.
Although IC reconstruction is fairly common at low redshifts, it remains relatively unexplored during the epoch of reionization (EoR; $z\gtrsim5$), in part due to the difficulties associated with modeling inhomogeneous cosmic radiation fields. And yet the EoR spans half of the observable Universe, offering unmatched potential to learn about astrophysics and cosmology.
Here we quantify how well upcoming galaxy and 21cm observations can constrain the ICs during the EoR. 
We develop \texttt{TILING} ({\bf T}omographic {\bf I}nference of {\bf L}inear {\bf I}Cs via {\bf N}etwork {\bf G}rafting): a novel, hybrid machine learning pipeline that first performs a point estimate of the ICs, which serves to improve the performance of a score-based diffusion network for generating posterior samples.
We train {\tt TILING} using mock galaxy maps at varying UV magnitude limits, as well as corresponding cosmic 21cm maps at varying noise levels and foreground ``wedge'' contamination.  
For fiducial observational survey choices, we obtain accurate IC reconstruction (e.g., with the posterior mean cross-correlation coefficients greater than 0.8 and error on the power spectra less than a few percent) at scales down to $\text{k}\lesssim0.2$ cMpc$^{-1}$.  Although 21cm interferometry helps in recovering the IC power spectra, most of the constraining power (especially for the Fourier phases of the ICs) comes from the galaxy maps, further highlighting the need for complementary observations when interpreting the cosmic 21cm signal.  
We showcase how {\tt TILING} can be used to recover modes of the 21cm power spectrum that were excised due to foreground contamination.  Additionally, our IC reconstruction framework can be used to: (i) guide follow-up observations of sub-volumes of interest; (ii) reconstruct the galaxy evolution and corresponding reionization morphology of specific volumes; and (iii) isolate the contribution to reionization by the vast majority of galaxies that will remain unobserved by optical/IR telescopes such as {\it JWST}.  We make our code publicly available.
 }

% \titlerunning 

\maketitle

% SECT 1: INTRODUCTION===========================================
% ===============================================================
% ===============================================================

\section{Introduction }
\label{sec:introduction}

\begin{figure*}[htb!]
    \centering
    \includegraphics[width=1.9\columnwidth]{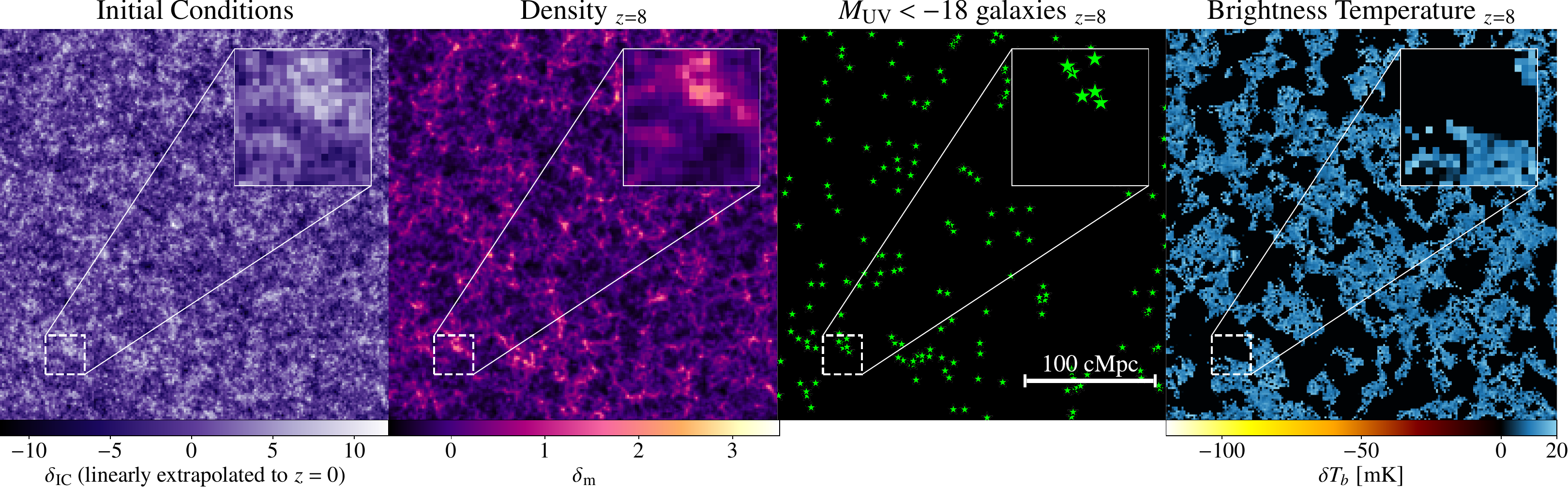}
    \hfill
    \caption{
     Illustration of the fields of interest. The matter ICs ({\it left panel}, showing $\delta_{\rm IC} \equiv \rho/\bar{\rho} - 1$ linearly extrapolated to $z=0$) are generated as a GRF, sampling a specified power spectrum. They are evolved through second-order Lagrangian perturbation theory to $z=8$, with the \textit{second panel} showing the filamentary structure of the large-scale density field. Dense collapsed structures (halos) host galaxies that emit radiation.  In the \textit{third panel} we show the bright ``observable'' galaxies with $M_{\rm UV}<-18$ as green stars. The combined radiation from all galaxies heats and ionizes the IGM, resulting in the 21cm brightness temperature field shown in the \textit{right panel}, where black regions indicate ionized gas. The last two fields are anti-correlated during the advanced EoR stages, since the rare observable galaxies are biased tracers of large-scale matter overdensities that also preferentially host the numerous faint neighboring galaxies that dominate the ionizing photon budget. Leveraging the complementarity of these two tracers may enhance IC reconstruction.
    }
    \label{fig:intro_plot}
\end{figure*}

In the standard cosmological model, early-Universe quantum fluctuations expanded to cosmological scales during inflation, seeding the matter over- and under-densities that define the initial conditions (ICs) of structure formation.  These ICs provide a clean probe of cosmology (e.g., see reviews in \citealp{Bartolo+2004, Weinberg+2013, Abdalla+2022, Cortes+2026}). 

The ICs of the matter fields are usually defined by their statistical properties. Observations of the cosmic microwave background (CMB) suggest that the ICs can be characterized (to high precision) as a homogeneous and isotropic Gaussian random field (GRF), whose statistics are fully specified by their power spectra (e.g., \citealt{Mather+1990, Smoot+1992, Ferreira+1998, Komatsu+2002, Komatsu+2003,Creminelli+2006, Bennett+2013, Planck+2016_XVII,  Planck+2020_IX, Planck+2020}).
Within this framework, the ICs of our observable Universe, or any sub-volume of it, correspond to a realization drawn from the underlying distribution that is determined by cosmology. However, our ability to infer full 3D maps of a target realization is severely limited if relying on CMB data alone. Putting aside secondary distortions which provide integral constraints,  the CMB effectively only maps 2D fluctuations in the photon-baryon fluid at the surface of last scattering, and probes relatively large scales (e.g., see reviews in \citealp{Scott+2010,Bucher+2015,Bianchini+2025, Dodelson+2020}).

Late-time probes such as galaxies and the intergalactic medium (IGM) can in principle provide additional constraints on the ICs and enable studies of specific cosmological volumes (e.g., \citealt{Monaco+1999,Wang+2014,Wang+2016,Tweed+2017, Modi+2018, Zhou+2024, Nguyen+2024, Parker+2025, Andrews+2026, Bayer+2026, Dong+2026}). However, such an approach comes with intrinsic difficulties. 
Following recombination, baryons can gravitationally condense and cool into dark matter halos, eventually forming the first stars, galaxies and black holes.  The radiation emitted by these sources heats and ionizes their surroundings, effectively coupling modes over cosmological scales. This combination of nonlinear gravitational dynamics and astrophysical/radiative processes makes it extremely challenging to infer the primordial ICs from late-time observations.

Despite these challenges, reconstructing the actual 3D realization of a given volume by effectively reversing time evolution offers unprecedented opportunities for astrophysics and cosmology.
Inferring the specific amplitude and phase of modes in a given region removes the need to marginalize over cosmic variance and phase uncertainties when interpreting complementary observations of the same volume, enabling  tighter parameter constraints
\citep{Sorce+2016, Darwish+2021}.
Furthermore, recovering the phase information makes higher-order correlations explicit, facilitating detections of primordial non-Gaussianity (PNG) signatures that would be inaccessible with the power spectrum \citep{Abazajian+2009, Andrews+2024, Floss+2024, Chen+2025_png, Bottema+2025, Andrews+2026_png}.

Driven by this potential, IC reconstruction at the field level (c.f., \citealp{Leclercq+2025} for a review of field-level inference) has become an active area of research, especially at low redshifts ($z<5$). 
Several approaches focus on point estimates rather than posterior inference for matter field reconstruction. These include optimization-based methods, such as the \texttt{TARDIS} algorithm using the Ly$\alpha$ forest (\citealp{Horowitz+2019, Horowitz+2021}, see also \citealp{Seljak+2017}), alongside other prominent point-estimate frameworks using alternative tracers \citep{Schmittfull+2017, Feng+2018, Modi+2021, Jindal+2023, Shallue+2023, Floss+2024, Wang+2024,  Bottema+2025}, while some narrow their scope strictly to late-time 2D density fields \citep{Ono+2024}.
Other efforts focus on Bayesian approaches that enable probabilistic reconstruction in the presence of modeling and observational uncertainties. Such large-scale structure (LSS) reconstructions have been widely applied by the \texttt{BORG} framework \citep{Jasche+2010, Jasche+2013, Jasche+2015, Jasche+2019, Jamieson+2023, Doeser+2024, Doeser+2025, Andrews+2026, Steinwandel+2026}. In these, galaxies are treated as biased tracers of the matter field, often assuming a Poisson likelihood when performing Bayesian inference (though see e.g. \citealp{Ata+2015}). 
Recent work has taken advantage of the efficiency of machine learning (ML) to accelerate or augment these pipelines 
\citep{Legin+2024, savchenko2024mean, Doeser+2024, Doeser+2025, Savchenko+2025}. 
Such reconstruction frameworks have been successfully applied 
to observational data, including SDSS-II \citep{Abazajian+2009}, BOSS/SDSS-III \citep{Eisenstein+2011, Dawson+2013}, the 2M++ catalog \citep{Lavaux+2011}, and the Quaia catalog \citep{Lang+2014, Meisner+2019, Delchambre+2023, StoreyFisher+2024} recovering the nearby LSS within projects such as \texttt{BORG, CLUES, SLOW} and \texttt{SIBELIUS} \citep{Lavaux+2019, Klypin+2003, Sorce+2016,  Gottloeber+2010, Dolag+2023, McAlpine+2022, Sawala+2022}. Beyond reconstructing the large-scale density field itself, these inferred initial conditions provide environmental constraints for high-resolution, constrained zoom-in simulations of specific objects as demonstrated for instance by the \texttt{CLUES}, \texttt{HESTIA} and \texttt{CLONE} projects \citep{Gottloeber+2010, Libeskind+2020, Sorce+2021, Sorce+2026} at $z\sim0$ and by \texttt{cosmosTNG} (\citealp{Byrohl+2025}) at $z\sim2$.

Compared to these lower redshifts, IC reconstruction at $z>5$ is relatively unexplored.  These high redshifts correspond to the Cosmic Dawn (CD) of the first galaxies and subsequent Epoch of Reionization (EoR).  During the CD and EoR, inhomogeneous cosmic radiation fields can have a dominant impact on observables.  Modeling such cosmological radiative transfer is computationally challenging, making IC reconstruction more numerically expensive than at lower redshifts.  Justifying such a theoretical effort is difficult given the historical dearth of high-quality observations probing the CD/EoR.

Luckily, in recent years we are starting to see a dramatic increase in CD/EoR data, both in terms of quality and quantity.  Broadband photometric surveys are pushing to deeper magnitudes and wider fields (see \citealp[e.g.,][Fig. 3]{Bagley+2026}). Narrow-band photometric programs have successfully achieved wide-field (tens of square degrees) coverage, resulting in large-scale galaxy maps targeting $\text{Ly}\alpha$ and providing crucial constraints on high-redshift populations well into the reionization era \citep[e.g.,][]{Konno+2014, Ouchi+2018, Inoue+2020, Aihara+2022, Umeda+2024}.
Space-based slit-spectroscopic campaigns with the James Webb Space Telescope (\textit{JWST})\footnote{\url{webbtelescope.org}} are confirming an increasing number of galaxies at these epochs \citep{Bagley2024, Donnan+2024, Finkelstein+2025, Eisenstein+2026}.
Concurrently, imaging with grism/slitless spectroscopy is helping bridge the traditional trade-off between survey area, depth, and redshift precision (\citealp{Kakiichi+2024, Meyer+2025}) and will continue to do so with the upcoming Nancy Grace Roman Space Telescope (\textit{RST})\footnote{\url{roman.gsfc.nasa.gov}}, which is poised to expand these efforts by mapping unprecedentedly wide fields  over thousands of square  degrees \citep{Perez+2023, Wold+2024, Roman+2025, Bagley+2026}.
Complementing these observations of resolved galaxies are upcoming line-intensity mapping (LIM) surveys, which aim to map the cumulative emission from unresolved galaxies over a wide area, using cross-correlation to remove foregrounds (e.g. \citealt{Bernal+2022, Chen+2025}). 
Finally, intensity mapping with the Square Kilometre Array (SKA)\footnote{\url{https://www.skao.int/en}} aims to directly map the neutral hydrogen (HI) using the 21cm line, providing the ultimate tomographic view of the first billion years (e.g., see \citealp{Liu+2020} for a review).

Using these upcoming EoR/CD observations to perform IC reconstruction offers 
additional scientific benefits, compared with studies at low redshifts.  Firstly, these epochs correspond to the bulk of our observable Universe, offering unmatched potential for future cosmic-variance-limited data. 
Secondly, instead of relying solely on costly blind surveys, 
IC maps can serve as spatial guides for complementary multi-wavelength observations.  
For example, targeted studies focusing on peaks in the matter field can search for signatures of early/large cosmic HII regions, active galactic nuclei (AGN)-driven quenching of neighboring galaxies, photo-heating feedback, and thermal fluctuations in the intergalactic medium (IGM). Alternatively, focusing on cosmic voids may increase the chances of finding first-generation, so-called Population III stars that might exist in relatively metal-poor regions even down to $z\sim 6-8$ (e.g. \citealt{Xu+2016, Venditti+2023, Zier+2025}). Lastly, having such a reconstruction would enable studies of the galaxy-IGM connection during the CD/EoR (see Sect. \ref{sec:conclusions} for further discussion), quantifying the role of more than 99\% of the galaxy population that is too faint to be seen by \textit{JWST} (e.g., \citealp{OShea+2015, Qin+2020}).  This unseen population of galaxies is expected to contribute the bulk of the ionizing photons responsible for reionization \citep{Qin+2024}, and yet there is no current or planned telescope capable of directly observing them.

To date, studies exploring IC reconstruction during the CD/EoR have been largely proof-of-concept, adopting idealized assumptions for both observations and simulations. \citet{Zhou+2024} estimated IC reconstruction from 21cm and CO line intensity maps of the EoR. Though pioneering in this field, their framework provided a point estimate that did not quantify the uncertainty of the reconstruction, nor did it account for foregrounds/systematics in the mock data.
On the other hand, \citet{Chen+2025} developed a framework to mitigate foreground/systematic contamination in mock 21cm observations, reconstructing the ICs in the process.  However, they use highly-idealized perturbative EoR models, do not account for important physical processes such as X-ray heating, and the quality of their IC reconstruction was limited by using only 21cm as a tracer field. It is worth noting that under these assumptions, their pipeline was able to recover the ICs reasonably well from different, more accurate simulations, which serves as an important out-of-distribution test.

In this work, we develop a Bayesian framework that uses galaxy and 21cm maps to reconstruct ICs during the EoR. Specifically, we create mock 3D galaxy maps at $z\sim8$ assuming different photometric limiting UV magnitudes, and compute corresponding 21cm maps for different integration times with the SKA, including foreground contamination.  We vary the ICs to make $2000$ realizations of these observables, quantifying reconstruction quality for different observational assumptions, as well as when using each tracer individually vs combined. To do this, we develop \texttt{TILING} ({\bf T}omographic {\bf I}nference of {\bf L}inear {\bf I}Cs via {\bf N}etwork {\bf G}rafting): a novel, two-stage hybrid machine learning pipeline that first performs a "best-guess" (point estimate) of the ICs, which then serves to improve the performance of a score-based diffusion network for generating posterior samples.
We make \texttt{TILING} publicly available at \url{https://github.com/nikos-triantafyllou/TILING}.

The paper is structured as follows. In Sect. \ref{sec:data} we describe how we compute the mock observations.
In Sect.  \ref{sec:methods} we introduce our \texttt{TILING} reconstruction pipeline.   
In Sect. \ref{sec:dependence_on_properties} we quantify the recovery performance under different observational assumptions and different combinations of tracers.  We demonstrate in Sect. \ref{sec:evolved_fields} how the recovered ICs can be used to re-simulate cosmic signals that are free from observational effects (for further such applications, see \ct{}).  We conclude in Sect. \ref{sec:conclusions}.  

Throughout we assume a standard $\Lambda \text{CDM}$ cosmology consistent with \cite{Planck_cosmo+2020}: $H_0=68 \, \text{km} \, \text{s}^{-1} \, \text{Mpc}^{-1}$, $\Omega_m = 0.31$, $\Omega_b = 0.049$, $\Omega_\Lambda =  0.69$, and $\sigma_8 = 0.81$. All lengths are quoted in comoving units and all Fourier transforms assume standard cosmological conventions, unless stated otherwise. When training ML models we make a standard 70/15/15 percent split for training/validation/testing. All inputs to ML models are standardized so that voxels of fields have zero mean and unit standard deviation.  Throughout, we consider the total matter field, assuming dark matter and baryons trace each other on the scales and redshifts of interest.

% SECT 2: DATA===================================================
% ===============================================================
% ===============================================================

\begin{figure*}[htb!]
    \centering
\includegraphics[width=1.9\columnwidth]{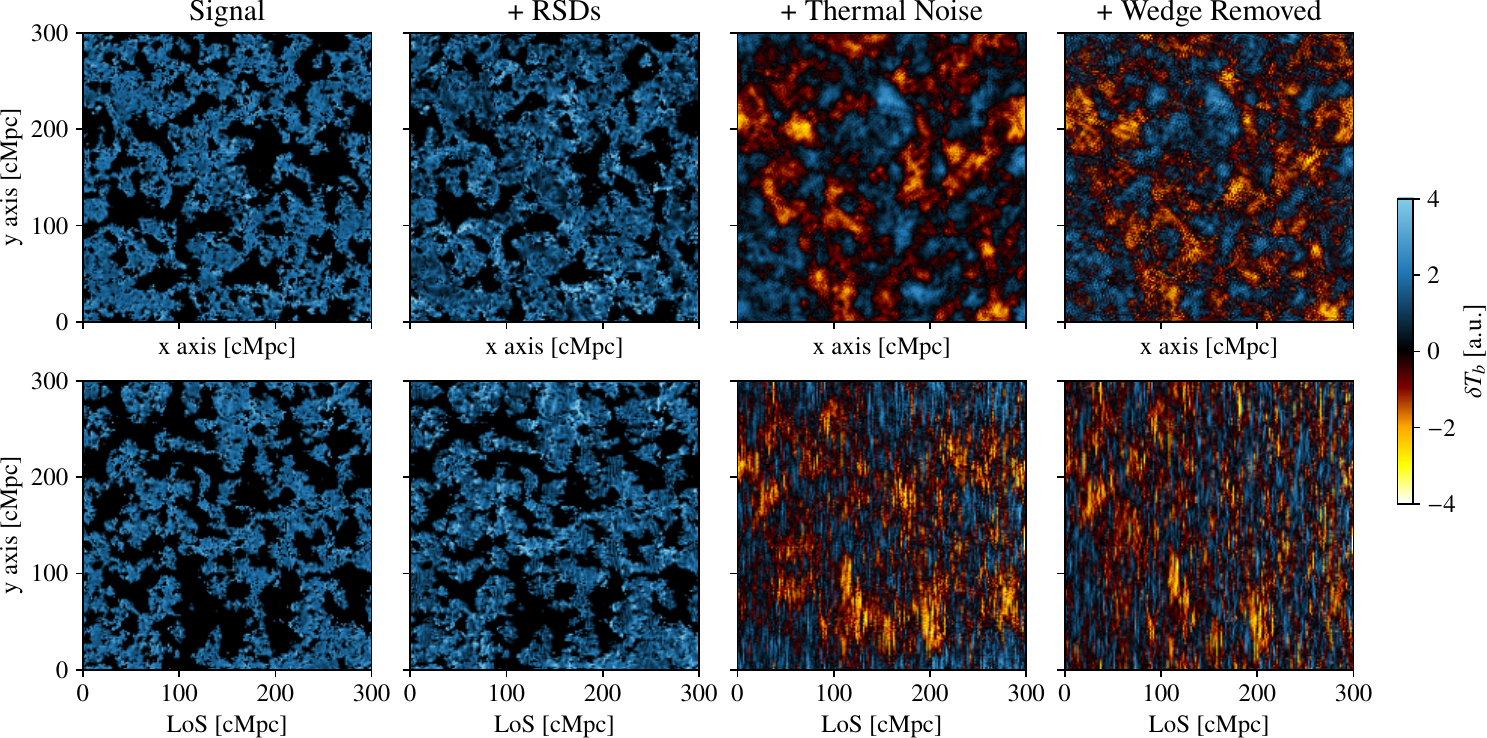}
    \hfill
    \caption{Impact of different observational effects on the 21cm signal.  The differential brightness temperature, $\delta T_b$ is shown
    %from SKA-low AA* (1000h) shown 
    for one-voxel-deep slices in the $\rm x-y$ plane (plane of the sky; \textit{top row}) and $\rm y-$line-of-sight (LoS) plane (\textit{bottom row}). From \textit{left} to \textit{right}, columns show: (i) the intrinsic cosmological signal; (ii) the intrinsic signal including redshift-space distortions (RSDs); (iii) the signal with added SKA-low AA* 1000h thermal noise; and (iv) the noisy signal after foreground wedge filtering.
    See text for details.
    For visual comparison, all fields have been normalized by their standard deviations ($5.71 \, \rm $, $5.06 \, \rm $, $332 \, \rm $ and $286 \, \rm mK$ respectively). The last two fields have undergone mean subtraction ($4.69\, \rm mK$) to mimic an interferometric observation.}
    \label{fig: 21cm_obs_effects}
\end{figure*}

\section{Simulating galaxy and 21cm maps} \label{sec:data}

As mentioned in the introduction we consider the two arguably most-promising observational probes of the EoR: galaxy counts and the 21cm signal (illustrated in the two right panels of Fig. \ref{fig:intro_plot}).
These probes are expected to be highly complementary.  Young stars in galaxies likely drive reionization.  Although most of the ionizing photons are likely provided by galaxies much too faint to be observed with current telescopes (e.g., \citealp{Qin+2024}), they still trace the same large-scale matter over-densities that host the bright, observed galaxies (albeit with different biases).  Conversely, the EoR 21cm signal is sourced by the HI outside of the cosmic HII regions, and so is correlated to large-scale matter underdensities that have a relative under-abundance of galaxies (for recent reviews, see e.g., \citealp{Pritchard+2012, Mesinger+2016}, see also, e.g., \citealp{Vrbanec+2020, Moriwaki+2024, Sam+2025, Hutter+2026}).

\subsection{Cosmological fields}
To simulate both fields, we use the public semi-numerical code \texttt{21cmFASTv4}\footnote{\url{https://github.com/21cmfast/21cmFAST/tree/2025-database-runs}} \citep{mesinger+2011, Murray+2020, Davies+2025}, which is intended for data-driven analyses of combined tracers such as IGM and galaxy observations (see Fig. \ref{fig:intro_plot} for example fields).
Starting from a realization of the ICs generated with \texttt{CLASS} \citep{Lesgourgues+2011, Blas+2011}, the density field is evolved to lower redshifts using second-order Lagrangian perturbation theory (2LPT; e.g.
\citealp{Scoccimarro+1998}).
Massive halos are identified using the \texttt{DexM} halo finder \citep{Mesinger+2007}, while sub-grid halos are constructed from conditional halo mass functions and merger trees (see \citealt{Davies+2025} for more details).
Galaxy properties are assigned to DM halos through physically-motivated, stochastic semi-empirical relations (discussed further in Sect. \ref{sect:galaxy_obs}).  For a given realization of galaxy properties, the corresponding inhomogeneous radiation fields are calculated using a combination of excursion-set photon counting and lightcone integration (see \citealt{mesinger+2011} for more details).

We run simulations with a box of side length $300 \rm \,cMpc$ and a resolution of $1.5\, \rm cMpc$. 
We use 3D coeval snapshots at $z=8$ for both galaxies and the 21cm signal.  This redshift is expected to be close to the EoR midpoint (e.g., \citealt{Qin+2024, Umeda+2024, Mason+2026}), which would roughly maximize the galaxy--21cm cross-correlation (e.g. \citealp{Sam+2025,Vrbanec+2020,Hutter+2026}).
We make 2000 realizations sampling different ICs (c.f., e.g.  \citealp{Doeser+2024,Savchenko+2025} for similar-sized training sets).  For this proof-of-concept we fix the astrophysical parameters to their default values (see Sect. \ref{sect:galaxy_obs} and \ref{sect:21cm_obs}), optimistically assuming that multi-wavelength EoR observations constrain the scaling relations below to reasonable accuracy (e.g., \citealp{Qin+2024}, \textcolor{blue}{Nikolić et al. in prep}).
Furthermore, although we account for redshift space distortions as discussed below, we ignore lightcone evolution over our $\rm 300\, cMpc$ box.  In future work we will relax these assumptions and explore more realistic forward-models matching specific surveys. 

\subsection{Galaxy maps}\label{sect:galaxy_obs}

As mentioned above, dark matter halos are populated with galaxies using physically-motivated empirical scaling relations (see \citealp{Davies+2025} and references therein for details regarding implementation and fiducial parameter choices). 
First, stellar masses are assigned to host halos by sampling from a given stellar-to-halo mass relation (SHMR) and corresponding scatter.  
Motivated by observations and hydrodynamic simulations (e.g., \citealp{Stefanon+2021, Harikane+2016, Kannan+2022, Pallottini+2022, DiCesare+2023}), we characterize the SHMR with a log-normal conditional distribution for the stellar mass, $M_\ast$, given a host halo mass, $M_h$: $p(\log M_\ast | \log M_h) = \mathcal{N}[\log M_\ast ~|~ \mu_\ast(M_h), \sigma_{M_{\ast}}]$, where the mean\footnote{For simplicity of notation, we use $\hat{\mu}$ to refer to the mean in linear space of log-normal distributions, given by $\hat{\mu}=\text{exp}\left(\mu+\sigma^2/2\right)$, where $\mu$ and $\sigma$ are the mean and standard deviations in log space.} follows a double-power law star formation efficiency (e.g., \citealt{Mirocha+17, Munshi+2021, Kar+2026}):
\begin{multline}\label{eq:acgstars}
    \hat{\mu}_* = f_{*,10} \left(
    \frac{(M_\mathrm{pivot}/10^{10}M_\odot)^{\alpha_{*}} + (M_\mathrm{pivot}/10^{10}M_\odot)^{\alpha_{*2}}}{(M_h/M_\mathrm{pivot})^{-\alpha_{*}} + (M_h/M_\mathrm{pivot})^{-\alpha_{*2}}} \right) \\
    \frac{M_h}{M_\odot}
    \exp \left(\frac{-M_\mathrm{turn}}{M_h} \right) ~ 
    \frac{\Omega_b}{\Omega_m}~.
\end{multline}
Here, $f_{*,10}= 0.05$ acts as a normalization of the SHMR at $10^{10} M_\odot$, while $(\alpha_*$, $\alpha_{*2})=(0.5,-0.61)$ are its low- and high-mass power-law slopes respectively, with $M_\text{pivot}=2.8\times 10^{11} \, M_\odot$ controlling their transition point.  $M_\text{turn}$ corresponds to the characteristic scale\footnote{In this work we include both atomic cooling galaxies and molecular cooling galaxies when calculating inhomogeneous radiation fields which shape the IGM properties (for more details see \citealp{Qin+2020, Munoz+2022};\ct{}).  To speed up the calculation, the contribution of the latter is summed over the entire cell without discretely sampling from the corresponding conditional halo mass function (see \citealp{Davies+2025}).  Nevertheless, the observable galaxies at $z=8$ are all hosted by relatively massive halos, orders of magnitude larger than $M_\text{turn}$, making our galaxy maps completely insensitive to this parameter.} below which inefficient gas cooling and/or feedback exponentially suppresses the abundance of halos hosting star-forming galaxies \citep{Silk+1998, Hopkins+2012, Sobacchi+2013, Xu+2016gal, Behroozi+2019, Nebrin+2023}, and is computed directly in the code based on the local radiation backgrounds. We set its lower limit to $M_\text{turn,low}=10^5\, M_\odot$, capturing the effect of stellar feedback.\footnote{Note that this uses the functionality of the specific version of \texttt{21cmFAST} used here. For more information the reader is referred to \url{https://21cmfast--748.org.readthedocs.build/en/748/M_TURN.html}.}
The standard deviation $\sigma _ *$ is taken to be a constant (0.3 dex), motivated by inference from {\it JWST} observations of ultra-violet luminosity functions (UVLFs) and angular correlation functions (\textcolor{blue}{Nikolić et al. in prep}).

Similarly, we relate the star formation rate (SFR) of each galaxy to its stellar mass with a conditional probability characterizing the star formation main sequence (SFMS) and associated scatter.  Again, motivated by hydrodynamic simulations we assume the SFMS follows a log-normal distribution $p_z(\log {\rm SFR} ~|~ \log M_\ast) = \mathcal{N}[\log {\rm SFR} ~|~ \mu_{\rm SFR}(M_\ast, z), \sigma_{{\rm SFR}}(M_\ast)]$, where the mean is defined following \citet{Park+2019}:
\begin{equation}\label{eq:sfrmean}
    \hat{\mu}_{\mathrm{SFR}} = \frac{M_*/M_\odot}{t_*H(z)} ~.
\end{equation}
Here $t_*$ is a free parameter corresponding to the characteristic star formation time-scale in units of the Hubble time $1/H(z)$ (which also scales with the mean dynamical time of halos during matter domination). We use the default value of $t_*=0.5$ (c.f. \citealt{Davies+2025}).  
The standard deviation decreases towards higher stellar masses as (e.g., \citealt{Davies+2025}, \textcolor{blue}{Nikolić et al. in prep}):
\begin{equation}\label{eq:sigmasfr}
    \sigma_\mathrm{SFR} = \mathrm{max} \left[ \sigma_\mathrm{SFR,lim},\sigma_\mathrm{SFR,idx}  \log \left( \frac{M_*}{10^{10}M_\odot} \right) + \sigma_\mathrm{SFR,lim} \right] ~,
\end{equation}
where $\sigma_\mathrm{SFR,lim}=0.19 \rm \, dex$ dictates the high-mass floor of the scatter, while $\sigma_\mathrm{SFR,idx}=-0.12$ is the power-law index as a function of stellar mass.

We compute the non-ionizing UV continuum ($\sim1500
   \text{\AA}$) from the SFR using the relation: 
\begin{equation}
    {\rm SFR}=\kappa_{\rm UV} \cdot L_{\rm UV}
\end{equation}
where $\kappa_{\rm UV}= 1.15\times 10^{-28} M_\odot \; \text{yr}^{-1}\; / \;\text{erg} \;\text{s}^{-1}\; \text{Hz}^{-1}$ is adopted from \cite{Sun+2016}, evaluated for continuous mode star formation with a Salpeter IMF and an evolving metallicity of $Z_*=10^{-0.15z} Z_\odot$. In the remainder of this work we will refer to this UV luminosity in units of absolute AB magnitudes calculated via the standard conversion (e.g., \citealp{Oke+1983}):
\begin{equation}
    \log_{10} \left( \frac{L_{\text{UV}}}{\text{erg s}^{-1} \text{Hz}^{-1}} \right) = 0.4 \times (51.63 - M_{\text{UV}}).
\end{equation}

Finally, we apply redshift space distortions (RSDs) using the simulated peculiar velocity field along the line of sight (LoS).  We first bin the galaxies to the native cell size to create a galaxy number density field.  We then perturb this field along the redshift axis, translating to the observed (redshift space) cell index $k^o$  (c.f., \citealp{Kaiser+1987})
\begin{equation}
    k^o = k^r + \frac{v_\parallel}{H(z)\,\Delta\chi},
\end{equation}
where $k^r$ is the comoving (real space) index, $v_\parallel$ is the line-of-sight velocity, $H(z)$ 
is the Hubble parameter evaluated at the redshift of cell $k^r$, and $\Delta\chi$ is the cell size in comoving coordinates. 
After mapping to redshift space, we distribute the galaxy number density linearly between two neighboring cells, following the standard cloud-in-cell weighing scheme (e.g., \citealp{Bagla+1997}).

\subsection{21cm maps}\label{sect:21cm_obs}

As described above, \texttt{21cmFASTv4} computes the IGM density evolution using 2LPT.  The temperature and ionization of each cell are tracked accounting for adiabatic and Compton heating/cooling, as well as heating and ionization from UV and X-ray radiation fields.  These are computed from the galaxy fields through a combination of approximate radiative transfer using excursion-set photon number counting, as well as integration along each cell's lightcone. Readers interested in further details are encouraged to consult \citet{Mesinger+2007, mesinger+2011, Davies+2025}.

From the IGM gas properties we can compute the 21cm brightness temperature field as
(e.g., \citealt{Furlanetto+2006}):
\begin{align}
\delta T_b & = \frac{T_s - T_\text{CMB}}{1+z} (1-e^{-\tau_{21}})
\\
\nonumber
&\stackrel{\text{$\tau_{21}\ll1$}}{\approx}  27\left(\frac{\Omega_b h^2}{0.023}\right)\left(\frac{0.15}{\Omega_m h^2}\right)^{1/2} x_{\rm HI}(1+\delta_{\rm m}) 
\\ 
\nonumber
& \;\;\;\;\;\;\;\;\; 
\left(\frac{1+z}{10}\right)^{1/2}
\frac{H(z)}{\mathrm{d}v_\parallel/\mathrm{d}r + H(z)}\left(1 - \frac{T_{\rm CMB}}{T_s}\right)
\,[\mathrm{mK}].
% \\ 
% \nonumber
% &\;\;\;\;\;\;\;\;\;\;
\label{eq:Tb}
\end{align}
Here, $\tau_{21}$ is the optical depth of the intervening gas, $T_s$ is the spin temperature of the underlying gas defined by the occupancy of the two hyperfine levels $n_{\uparrow\uparrow}/n_{\uparrow\downarrow} = 3 e^{-0.068\,\mathrm{K}/T_s}$, 
$\delta_{\rm m}\equiv \rho/\bar{\rho} - 1$ is the Eulerian (evolved) matter 
overdensity, $x_{\rm HI}$ is the neutral hydrogen fraction and $\mathrm{d}v_\parallel/\mathrm{d}r$ 
is the comoving gradient of the line-of-sight velocity field.  We note that {\tt 21cmFAST} computes the exact expression in the first row, while the approximation in the second row serves to gain physical insight about the relevant IGM properties.  
RSDs are implemented analogously to the galaxy field (see also \citealp{Mao+2012, Jensen+2013,Greig+2018}). The effect of RSDs on the intrinsic signal can be seen by comparing the first and second columns of Fig. \ref{fig: 21cm_obs_effects}.
We use a modified version of the post-processing package \texttt{tuesday}\footnote{\url{https://github.com/21cmfast/tuesday/tree/better_lc_noise}} (\citealp{Breitman+2025}) to add realizations of thermal noise to the cosmic signal, assuming the AA* antenna layout of SKA-low \citep{Sridhar+2023}.
\begin{figure*}
    \centering
    \includegraphics[width=1.95\columnwidth]{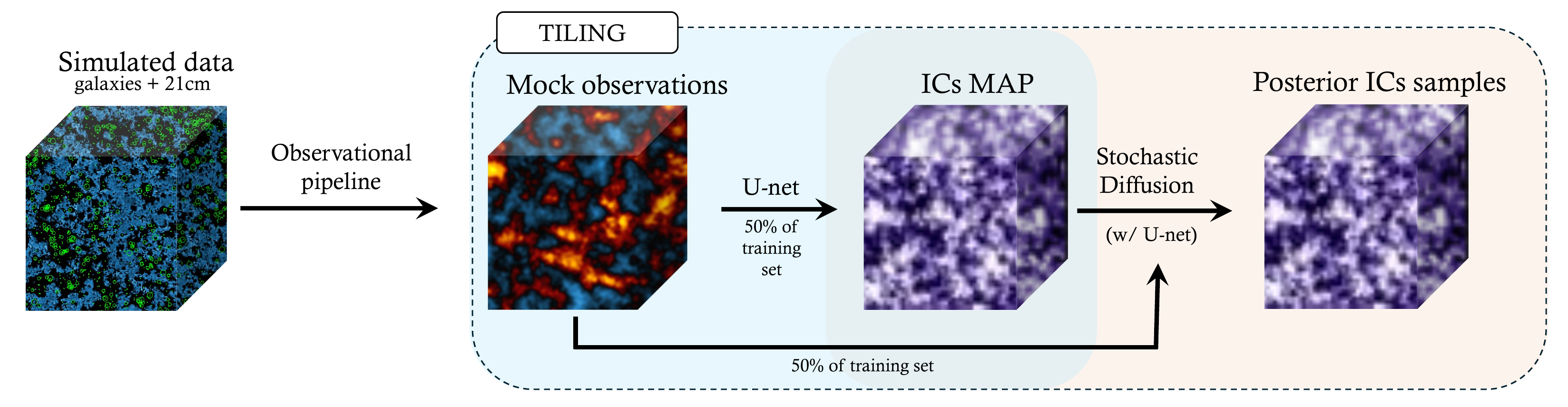}
    \includegraphics[width=1.55\columnwidth]{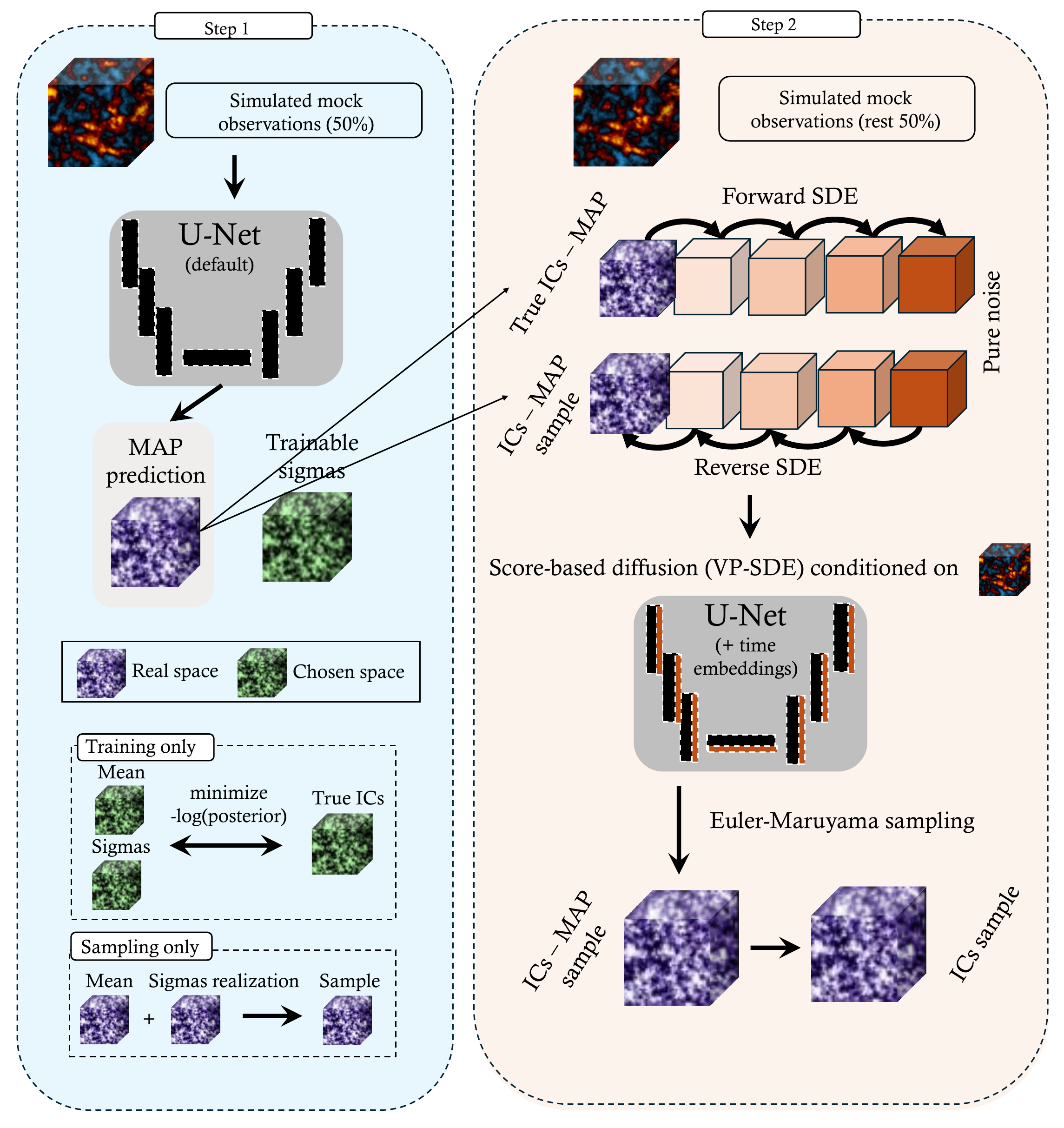}
    \hfill
    \caption{Schematic of our inference pipeline. {\it Top panel}: The two simulated fields (galaxies represented as green circles on top of the 21cm brightness temperature field, see also Fig. \ref{fig:intro_plot}) are passed through an observational pipeline (see Fig. \ref{fig: 21cm_obs_effects}) and then to a hybrid NN model, {\tt TILING}, to predict the ICs. {\tt TILING} consists of two parts: the first uses 50\% of the training data to predict a maximum a posteriori (MAP) estimate of the ICs, which is used together with the remaining 50\% of the training set by the second NN to generate posterior samples of the ICs. {\it Bottom left panel}: the first part of {\tt TILING} (based on \citealp{Savchenko+2025}). The mock data are passed through a U-net (architecture is shown in Appendix \ref{appdx:NN_architecture}) to predict a mean IC field in real space. At the same time the sigmas of the uncertainty are also predicted in Hartley space. During training, the -log(posterior) is minimized as explained in Sect. \ref{sec:methods}, while, during sampling, a realization of the sigmas is drawn and added into the mean prediction in real space to create a sample. We only use this sigma for comparison purposes in Appendix \ref{appdx:NN_architecture}, and potentially to stabilize the training as it is quantifying aleatoric homoscedastic uncertainty (e.g., \citealp{Kendall+2017}). {\it Bottom right panel}: The second part of {\tt TILING}. A score-based diffusion model is trained (as in \citealp{Song+2020}) to denoise a field from the reverse variance preserving (VP) stochastic differential equation (SDE). The target field is the residual field of the true ICs and the MAP prediction from the first part. The same U-net is being utilized but with the addition of a time embedding (see Appendix \ref{appdx:NN_architecture}) and being conditioned on the mock data (apart from the noisy field of the diffusion). The residual is sampled through an Euler-Maruyama scheme, to which the MAP is added for the final IC prediction.}
    \label{fig: gaussian_schematic}
\end{figure*}
Specifically, for each frequency slice $\nu_i$, for both real and imaginary components, the root-mean-square (RMS) of the thermal noise per baseline per snapshot is computed by \texttt{tuesday} as (e.g., \citealt{Giri+2018})
\begin{equation}
    \sigma_{uv} = \frac{T_{\rm sys}\,\Omega_{\rm beam}/\Omega_{\rm pix}}
    {\sqrt{2\,\Delta\nu\,t_{\rm int}}},
    \label{eq:sigma_uv}
\end{equation}
where $T_{\rm sys}= T_{\rm sky} + T_{\rm receiver} =[60\,(\nu_i/360\, \text{MHz})^{-2.55} + 100]\,\rm K$ is the system temperature, $\Omega_{\rm beam}=(c/\nu_i)^2/A_{\rm eff} \rm \, sr$ is the effective
beam area (with $A_{\rm eff}=962\, \rm m^2$), $\Omega_{\rm pix} = [\Delta\chi/d_c(z)]^2 \, \rm sr$ is the solid angle 
of a single pixel with $d_c$ being the comoving distance to redshift $z$, $\Delta\nu$ is the channel bandwidth (corresponding to the simulation's resolution of $\Delta\chi = 1.5\,\rm cMpc$), and 
$t_{\rm int}=60$ s is the integration time per snapshot. All values are chosen as the default in \texttt{tuesday} and are similar to \cite{Prelogovic+2022}.
Throughout, we adopt a 6h per day observation window, allowing the sky to rotate naturally and the baselines to populate the $uv$ plane via rotational synthesis. 
Then the total RMS noise 
per $uv$ cell is obtained by dividing by the square root of the number 
of baselines sampling that cell accumulated over multiple snapshots $N_{\rm snapshots}$ as:
\begin{equation}
    \sigma_{uv}^{\rm total} = \frac{\sigma_{uv}}{\sqrt{N_{uv}\,N_{\rm snapshots}}},
\end{equation}
where $N_{uv}$ is the number of baselines per $uv$ cell per snapshot. Cells with no 
baseline coverage are set to zero. A noise realization is then drawn in 
$uv$ space by sampling a complex Gaussian with RMS $\sigma_{uv}^{\rm total}$
(respecting Hermitian constraints to ensure a real-valued output), and then added to the signal. We also subtract the mean (i.e., DC mode) of each frequency slice, mimicking interferometric observations. 

For each 2D gridded signal in $uv$ space we weight by the inverse variance of the noise and Fourier transform back to configuration space (i.e., we use `natural' weighting). 
The effect of the noise addition under this weighting can be seen by comparing the second and third columns of Fig. \ref{fig: 21cm_obs_effects}. This noise injection yields a mean-subtracted, smoothed field that erases small-scale structural information.
For a further discussion on weighting schemes, the reader is referred to Appendix \ref{sect:ska_obs}.

Finally, foreground-contaminated modes 
are removed by applying a wedge filter in $(k_\parallel, k_\perp)$ space, zeroing all modes satisfying
(e.g., \citealp{Liu+2014,Murray+2018, Morales+2012, Prelogovic+2022}):
\begin{equation}
    k_{\parallel} < \frac{H(z) d_c(z)}{c(1+z)} \text{sin}\theta \cdot k_{\perp} + b~.
\end{equation}
Here, $d_c$ is the comoving distance to redshift $z$, $\theta$ is a characteristic width related to the beam of the interferometer, and $b$ is 
a buffer term, which we omit for this analysis. We assume a $\theta=9\, \rm deg$, which is between the horizon and the SKA FoV of $\sim1.7 \, \rm deg$.
This choice is somewhat optimistic, though reasonable considering state-of-the-art wedge reconstruction schemes (e.g., \citealt{Gagnon+2021, Kennedy+2024, Bianco+2024, Chen+2025, Mertens+2025, Munshi+2025}).
As seen by comparing the third and fourth columns of Fig. \ref{fig: 21cm_obs_effects}, this wedge excision results in the most severe degradation of the 21cm signal topology.

% SECT 3: METHODS================================================
% ===============================================================
% ===============================================================

\section{Inferring ICs from tracer fields during the EoR}\label{sec:methods}

Our goal is to use the mock 3D maps of galaxy number counts and 21cm brightness temperature discussed in the previous section to infer the 3D initial conditions in a Bayesian way:
\begin{equation}
    p(\boldsymbol\theta|\boldsymbol d) \propto p(\boldsymbol d|\boldsymbol \theta)\cdot p(\boldsymbol \theta)
\end{equation}
where the data vector $\boldsymbol d$ corresponds to the galaxy and/or $\delta T_b$ fields and $\boldsymbol \theta$ are the ICs.  
Inherently, this is a very challenging task due to: (i) the dimensionality of $\boldsymbol{\theta}$ (e.g., for our set up, $\boldsymbol{\theta}$ has a dimensionality of 200$^3$=$8\times10^6$); as well as (ii) the lack of an analytically-tractable likelihood for such highly non-Gaussian fields like galaxies and 21cm. 
Machine learning and neural networks (NNs) are starting to be used to tackle (i) and (ii) individually (e.g., \citealp{Floss+2024,Prelogovic+23, Cranmer+2020}); however, scaling them to accurately deal with both together remains a challenge.

To address this, we introduce a novel two-step machine learning pipeline. The first stage  estimates the maximum a posteriori (MAP; e.g., \citealt{Savchenko+2025}) of $\boldsymbol{\theta}$, while the second stage uses stochastic diffusion (e.g., \citealt{Legin+2024}) to improve this estimate and provide posterior samples.  As quantified in Appendix \ref{appdx:NN_architecture}, this hybrid approach results in more accurate reconstruction than using each network by itself.  We describe these in more detail below, and provide a detailed schematic in Fig. \ref{fig: gaussian_schematic}.

\subsection{Step 1: Making a ballpark point estimate}

For the first stage, we adapt the neural Gaussian posterior estimation (NGPE) framework introduced by \cite{Savchenko+2025}, as illustrated in the bottom left panel of Fig. \ref{fig: gaussian_schematic}.  We describe in turn the components of this NGPE.

\subsubsection{Prior}
We assume the ICs are a homogeneous and isotropic Gaussian random field,  characterized by a power spectrum $\rm P(k)$ ($\rm k=|\boldsymbol{k}|$); i.e., each 3D Fourier mode is statistically independent, with an amplitude drawn from a  Gaussian distribution with variance $P(|\boldsymbol{k}|)$, and a phase uniformly distributed over $(0, 2\pi)$.  The prior in configuration space can then be written as:
\begin{equation}
    p(\boldsymbol \theta) = \frac{\rm{exp}\left( -\frac{1}{2} \boldsymbol{\theta}^T (\boldsymbol{C}^{prior})^{-1}\boldsymbol{\theta} \right)}{\sqrt{(2\pi)^D \rm{det}(\boldsymbol{C}^{prior})}}.
\end{equation}
The covariance is $\boldsymbol{C}^{\rm prior}=\boldsymbol{\mathcal{F}}^{-1}\boldsymbol{D}^{\rm prior}(\text{k})\boldsymbol{\mathcal{F}}$,
where $\boldsymbol{\mathcal{F}}$ is the matrix representation of the unitary discrete Fourier transform (DFT), and $\boldsymbol{D}^{\rm prior}(\text{k})$ a diagonal matrix with diagonal values $\rm P(k)$.

\subsubsection{Likelihood}

Our NN has as input a $200^3$ galaxy and/or 21cm map $\boldsymbol d$, and as output an estimate of the corresponding $200^3$ map of ICs, ${\boldsymbol\theta}^\text{L}_\text{NN}(\boldsymbol d)$, together with the estimate's uncertainty through a covariance matrix $\boldsymbol C_\text{NN}^\text{L}$ (independent of $\boldsymbol d$).\footnote{Even though we use this first NN only as a point estimate, we include a variance prediction since it has been shown to improve training, as it quantifies aleatoric homoscedastic uncertainty (e.g., \citealp{Kendall+2017, Kendall2017b}). We also use it for comparison purposes in Appendix~\ref{appdx:NN_architecture}.}
Intuitively, the NN acts as a last step of an effective data reduction pipeline. It transforms the data $\boldsymbol d$ into IC space, producing an effective estimate of the ICs, ${\boldsymbol\theta}^\text{L}_\text{NN}(\boldsymbol d)$, with effective uncertainty given by $\boldsymbol C^\text{L}_\text{NN}$. 
 We take this to define our likelihood ansatz: a multivariate Gaussian in IC space, with mean ${\boldsymbol\theta}^L_\text{NN}(\boldsymbol d)$ and covariance $\boldsymbol C^\text{L}_\text{NN}$. That is:
\begin{equation}       
    p[{\boldsymbol{\theta}^\text{L}_\text{NN}(\boldsymbol{d})}|\boldsymbol{\theta}] =  \frac{ \rm{exp}\left\{ -\frac{1}{2} (\boldsymbol{\theta}-{\boldsymbol{\theta}}^{L}_\text{NN}(\boldsymbol{d}))^T(\boldsymbol{C}_\text{NN}
    ^L)^{-1} (\boldsymbol{\theta}-{\boldsymbol{\theta}}^{L}_\text{NN}(\boldsymbol{d}))\right\}}{\sqrt{(2\pi)^D\rm{det}(\boldsymbol{C}_\text{NN}^L)}}
\end{equation}
Because this Gaussian is defined over $\boldsymbol\theta$, its mode for fixed $\boldsymbol d$ is exactly ${\boldsymbol\theta}^\text{L}_{\text{NN}}(\boldsymbol d)$; due to Gaussianity, this makes ${\boldsymbol\theta}^\text{L}_\text{NN}(\boldsymbol d)$ the maximum likelihood estimate (MLE) under this ansatz, before any prior information is included.

\subsubsection{Posterior}

Multiplying the likelihood and the prior, results in the following posterior $p[\boldsymbol{\theta}|\boldsymbol{\theta}_\text{NN}^\text{L}(\boldsymbol{d})] = p[\boldsymbol{\theta}|\boldsymbol{d}, \text{NN}] $ and so
\begin{equation}\label{eq:posterior}
    \hat{p}(\boldsymbol{\theta}|\boldsymbol{d}) = \frac{\rm{exp}\left\{ - \frac{1}{2} (\boldsymbol{\theta}-\boldsymbol{{\theta}}^{MAP}_\text{NN}(\boldsymbol{d}))^T(\boldsymbol{C}^p_\text{NN})^{-1} (\boldsymbol{\theta}-\boldsymbol{{\theta}}^{MAP}_\text{NN}(\boldsymbol{d}))\right\}}{\sqrt{(2\pi)^D\rm{det}(\boldsymbol{C}_\text{NN}^p)}},
\end{equation}
where the hat indicates an approximation of the posterior via a NN.  
Here, we merged the covariances of the prior and the likelihood into the learnable covariance of the posterior $\rm (\boldsymbol{C}_\text{NN}^p)^{-1} = (\boldsymbol{C}_\text{NN}^L)^{-1} + (\boldsymbol{C}^{prior})^{-1}$. This makes explicit that the prior's information is fully retained in the posterior, rather than discarded; non-Gaussian priors would combine differently (e.g., \textcolor{blue}{Flitter et al. in prep}). Similarly, $\boldsymbol{\theta}_\text{NN}^\text{MAP}$ here is the mean of the posterior that is also the MAP due to Gaussianity.

\subsubsection{Training}
The training objective is to make $\hat p(\boldsymbol\theta|\boldsymbol d)$ approximate the true posterior $p(\boldsymbol\theta|\boldsymbol d)$ as closely as possible by optimizing the NN's parameters $\textbf{w}$ via stochastic gradient descent and backpropagation. To do so, we minimize a distance between the two distributions, e.g., the Kullback--Leibler (KL) divergence:
\begin{align}
    D_{\mathrm{KL}}(p \parallel \hat p) &\equiv  \int p(\boldsymbol{\theta} | \boldsymbol{d}) \log \left( \frac{p(\boldsymbol{\theta} | \boldsymbol{d})}{\hat p(\boldsymbol{\theta} | \boldsymbol{d})} \right) d\theta
    \\&=\int p(\boldsymbol{\theta} | \boldsymbol{d}) \log p(\boldsymbol{\theta} | \boldsymbol{d}) \, d\theta - \int p(\boldsymbol{\theta} | \boldsymbol{d}) \log \hat p(\boldsymbol{\theta} | \boldsymbol{d}) \, d\theta \nonumber 
\end{align}
This needs to be true for every possible data $\boldsymbol{d}$ (from the prior predictive distribution) and hence
\begin{align}
    & \int\!\!\int p(\boldsymbol\theta,\boldsymbol d) \log p(\boldsymbol\theta|\boldsymbol d) \, d\boldsymbol\theta \, d\boldsymbol d - \int\!\!\int p(\boldsymbol\theta,\boldsymbol d) \log \hat p(\boldsymbol\theta|\boldsymbol d) \, d\boldsymbol\theta \, d\boldsymbol d \nonumber \\
    &= I - \int\!\!\int p(\boldsymbol\theta,\boldsymbol d) \log \hat p(\boldsymbol\theta|\boldsymbol d) \, d\boldsymbol\theta \, d\boldsymbol d,
\end{align}
where $I$ is a constant with respect to the NN's parameters and can be dropped. Crucially, $p(\boldsymbol\theta,\boldsymbol d) = p(\boldsymbol\theta)\,p(\boldsymbol d|\boldsymbol\theta)$ is precisely the joint distribution our simulator samples: we draw $\boldsymbol\theta$ from the prior, then simulate $\boldsymbol d$ given $\boldsymbol\theta$. This means the remaining double integral can be estimated directly using training pairs $(\boldsymbol\theta_i,\boldsymbol d_i)$ drawn from the simulator, via a Monte Carlo average:
\begin{equation}
    \int\!\!\int p(\boldsymbol\theta,\boldsymbol d) \log \hat p(\boldsymbol\theta|\boldsymbol d) \, d\boldsymbol\theta \, d\boldsymbol d \approx \frac{1}{N}\sum_{i=1}^N \log \hat p(\boldsymbol\theta_i|\boldsymbol d_i).
\end{equation}
Minimizing the KL divergence hence leads to the per instance loss\footnote{We also investigated an alternative, multi-scale loss function incorporating power spectrum $\rm P({\rm k})$ weighting; however, we find that this spectral weighting has a negligible effect on training convergence and prediction quality, leading us to retain the simpler formulation.}:
\begin{align}
\begin{aligned}
    \mathcal{L}{\rm oss}(\textbf{w}) &= -{\rm log}\left\{ p(\boldsymbol{\theta}_{i
    }|\boldsymbol{d_i}) \right\} = \\
&=  \frac{1}{2} (\boldsymbol{\theta}_i-\boldsymbol{{\theta}}^{\text{MAP}}_\text{NN}(\boldsymbol{d_i}))^\text{T}({\boldsymbol{C}}^\text{p}_\text{NN})^{-1} (\boldsymbol{\theta}_{i}-\boldsymbol{{\theta}}^{\text{MAP}}_\text{NN}(\boldsymbol{d}_i)) \\
&~~~~~~~~~~~+ \frac{D}{2}{\rm log}(2\pi) + {\rm tr}~{\rm log}(\boldsymbol{C}_\text{NN}^\text{p}),
\end{aligned}
\end{align}
that is averaged over each batch during training.

We assume a diagonal covariance in Hartley space under symmetric Fourier transformations. Hence, the loss is practically evaluated in Hartley space during training. We utilize a modified U-net architecture to predict $\rm \boldsymbol{{\theta}}^{MAP}_\text{NN}$ and $\rm \boldsymbol{C}_\text{NN}^p$ based on \cite{Savchenko+2025} and \cite{Jamieson+2023} from the \texttt{map2map} repository\footnote{\url{https://github.com/eelregit/map2map}} (see Appendix \ref{appdx:NN_architecture} for details).

We train this part of the hybrid model with $50\%$ of the available dataset %\footnote{We also tested using the $25\%$ or the $75\%$ of the dataset but found no notable change in the metrics considered (see Sect. \ref{sec:dependence_on_properties} for definitions). This further indicates that the choice of 2000 simulations is sufficient for the purposes of this work. Nevertheless, sampling astrophysical uncertainties would likely require a larger training set.} 
for up to 500 epochs, utilizing an early stopping patience of 40 epochs based on validation loss. We also investigated whether alternating training between the MAP and covariance parameters would yield more accurate results, but found no significant improvement.

\subsection{Step 2: Predicting posterior samples with score-based diffusion}

Given that the first part of  our hybrid model produces an estimate of the MAP assuming a Gaussian diagonal covariance, the goal of this second step is to predict the distribution of the residuals, $\mathbf{x} = \boldsymbol{\theta}_\text{NN}^{\rm MAP}(\boldsymbol{d}) - \boldsymbol{\theta}_{\rm true}$ (where we renamed $\boldsymbol{\theta}_i$ to $\boldsymbol{\theta}_\text{true}$ for clarity) using the remaining $50\%$ of the available dataset (illustrated in the top panel of Fig. \ref{fig: gaussian_schematic}).\footnote{No improvement was seen in the metrics considered (see Sect. \ref{sec:dependence_on_properties} for definitions) by splitting the dataset differently between the two parts of \texttt{TILING} (a 25/75 and a 75/25 split were tested).}

% \footnote{No improvement was seen in the metrics considered (see Sect. \ref{sec:dependence_on_properties}) by splitting the dataset differently between the two parts of \texttt{TILING} (a 25/75 and a 75/25 split were tested).  We also did not see appreciable differences either even when considering the first part of \texttt{TILING}, indicating that our choice of 2000 simulations is sufficient. To this end we also expect the modes that can be expressed in smaller subvolumes to be well represented within the 2000 simulations, due to our simplifications for the astrophysical parameters, and due to ergodicity. The sufficiency statement is further enhanced by the similar number of total simulations in similar works (e.g., \citealp{Doeser+2024,Savchenko+2025})}

To do so, we use state-of-the-art score-based diffusion models. Specifically, we use the variance preserving (VP) stochastic differential equation (SDE) version of \cite{Song+2020}, which is a generalization of the discrete denoising diffusion probabilistic models (DDPMs) of \cite{ho2020denoising, sohl2015deep}. Other flavors of SDEs were tested, including VE-SDE and sub-VP SDE, but they resulted in similar or worse results.

In the DDPM formulation, a forward process adds noise to the field $\mathbf{x}$ of interest over $T$ discrete  timesteps through the Markov chain:
\begin{equation}
    \mathbf{x}_t \stackrel{\mathrm{def}}{=} \sqrt{1-\beta_t} \mathbf{x}_{t-1} + \sqrt{\beta_t}\mathbf{z}
\end{equation}
where $\mathbf{z}\sim\mathcal{N}(\mathbf{0},\mathbf{I})$, and $\beta_t$ defines a variance schedule.

We use the generalization of this discrete process by taking the limit as the number of steps $T \to \infty$ (and $\Delta t \to 0$). This transforms the noise injection into a VP-SDE (the reader is referred to \citealp{Song+2020} for a more extensive analysis):
\begin{equation}
\label{eq:main_sde}
    d\mathbf{x} = -\frac{1}{2}\beta(t)\mathbf{x} dt + \sqrt{\beta(t)} d\mathbf{w},
\end{equation}
with a linear schedule
\begin{equation}
    \beta(t) = \bar{\beta}_{\min} + t(\bar{\beta}_{\max} - \bar{\beta}_{\min}) \quad \text{for} \quad t \in [0, 1],
\end{equation}
where $d\mathbf{w}$ is a Wiener process characterizing the random noise. 
Given Eq. \ref{eq:main_sde}, the field at time $t$ is given by:
\begin{equation}\label{eq:trans_kernel}
\begin{aligned}
    p(\mathbf{x}_t \mid \mathbf{x}_0)  &= \mathcal{N} \Bigl( \mathbf{x}_t; e^{-\frac{1}{4} t^2 (\bar{\beta}_{\max} - \bar{\beta}_{\min}) - \frac{1}{2} t \bar{\beta}_{\min}} \mathbf{x}_0, \\ 
    &\quad \mathbf{I} - \mathbf{I} e^{-\frac{1}{2} t^2 (\bar{\beta}_{\max} - \bar{\beta}_{\min}) - t \bar{\beta}_{\min}} \Bigr), \quad t \in [0, 1].
\end{aligned}
\end{equation}
The backward process of Eq. \ref{eq:main_sde} is also a diffusion process \citep{Anderson1982-ny}, 
\begin{equation}
    \mathrm{d}\mathbf{x} = \left[ -\frac{1}{2}\beta(t)\mathbf{x}  - {\beta(t)}  \nabla_{\mathbf{x}} \log p_t(\mathbf{x}) \right] \mathrm{d}t + \sqrt{\beta(t)}  \mathrm{d}\bar{\mathbf{w}},
\end{equation}
in which the only unknown is the score $\nabla_{\mathbf{x}} \log p_t(\mathbf{x})$, which we extend to be the conditional score $\nabla_{\mathbf{x}} \log p_t(\mathbf{x}|\boldsymbol{d}
)$ to include our observational constraints.

When trained on forward models generated from prior samples, every call to {\tt TILING} conditioned on data should result in a posterior sample, as the network learns the conditional score (analogous to other forms of simulation-based inference in which NNs are used for density estimation; e.g., \citealp{Cranmer+2020}).  In practice, {\tt TILING} estimates the score through denoising score matching: 
 for a uniformly sampled $t\in [\epsilon,1]$ in Eq. \ref{eq:trans_kernel} the following objective function is minimized (Eq. 7 of \citealp{Song+2020})
\begin{equation} 
\mathcal{L}{\rm oss}(\textbf{w})=  \lambda(t) \mathbb{E} \left[ \left\| \mathbf{s}_\textbf{w}({\mathbf{x}_t}, t, \boldsymbol{d}) + \frac{\mathbf{z}}{\sigma(t)} \right\|^2 \right] ,
\end{equation}
where $\sigma(t)$ is the standard deviation from Eq. \ref{eq:trans_kernel} and $\lambda(t)$ is a weighting function which we commonly set to $\sigma(t)^2$ which transforms the score into a noise prediction objective loss. This reparameterization ensures numerical stability, especially for half-precision floating-point operations, as the pure score scales as $1/\sigma(t)$, which can exceed the dynamic range as $\sigma(t)\to 0$.
The lower limit $\epsilon=10^{-5}$ is commonly added for numerical stability.

For the sampling, we use the common Euler-Maruyama method (e.g., \citealp{Legin+2024}):
\begin{equation}
x_{t+\Delta t} = x_t 
+ \left( -\frac{1}{2}\beta(t)\mathbf{x}  - \beta(t)\nabla_{x_t} \log p_t(x_t |\boldsymbol{d}) \right)\Delta t
+ \sqrt{\beta(t)} z_t \sqrt{-\Delta t},
\end{equation}
where $\Delta t = -1/N$ is the step size, $N=1000$ is the number of steps, and 
$z_t$ is sampled from a standard normal distribution with the same dimensions as $x_t$.

We use the same network as in the first part of the hybrid model but, 
following common approaches, we add a Fourier time embedding at every convolution before the activation functions. We use the same embedding as in \cite{Legin+2024} while reducing the last dimension of the embedding linear layer by 4 to inject it into our U-net and using the default initialization of \texttt{PyTorch} (\citealp{Paszke+2019}\footnote{\url{https://pytorch.org/}}).  We use an exponential moving average (EMA), that is, the weights of the NN are calculated as $\rm \theta_{EMA}=\alpha \theta_{EMA}+ (1-\alpha)\theta_{now}$ at each epoch, where $\alpha$ is the decay parameter set at $\alpha=0.999$ as suggested by \cite{Song+2020} for a VP-SDE, like in our case. We trained the diffusion model for up to 300 epochs and selected the model that maximized $\rm k_{08}$ (or $\rm k_{04}$ in setups where $\rm k_{08}$ was not defined) while keeping the transfer function below 10\% when evaluated on a stochastically sampled validation subset (see the next section for details on these metrics). Everything was run on 4 NVIDIA A100-SXM-64GB GPUs. For more details, the reader is referred to Appendix \ref{appdx:NN_architecture}.

Lastly, the integration of the MAP estimate from the first stage into the diffusion is not trivial and can be a critical design choice. We investigated three conditioning schemes:
\begin{itemize}
    \item Multi-channel input: concatenating the NGPE MAP prediction as an extra input channel into the U-net.
    \item Time-inspired embedded injection: adding the NGPE prediction to every convolutional layer (normalizing it by the number of additions).
    \item Residual prediction: training the diffusion model to predict the residual $\mathbf{x} = {\boldsymbol{\theta}}_\text{NN}^\text{MAP}(\mathbf{d}) - \boldsymbol{\theta}_\text{true}$.
\end{itemize}
The latter consistently yielded the highest reconstruction fidelity in terms of the metrics considered (see Sect. \ref{sec:dependence_on_properties}) and was chosen as the final architecture.

% SECT 4: RESULTS================================================
% ===============================================================
% ===============================================================

\section{Dependence of recovery on survey properties}\label{sec:dependence_on_properties}

In the previous section, we outlined the methods used to recover the $200^3$ voxel fields of ICs  from  $300^3\,\rm cMpc^3$ galaxy and 21cm mock observations at $z=8$. Here, we use the test set to compare the posterior samples of the reconstructed fields with their corresponding ground truths across different observational scenarios.
Aside from visual inspections, we employ two commonly used statistical metrics for such fields: the (1D) power spectrum and the cross-correlation coefficient (CCC).

Given a realization of $\boldsymbol{\delta}_{\rm IC}({\bf x})$ in configuration space, the (1D) power spectrum is defined via its 3D Fourier transform (e.g., \citealp{Sirko+2005}) 
$\boldsymbol{\delta}_{\rm IC}(\boldsymbol{k}) \equiv \delta_{\rm IC, \boldsymbol{k}} $ as:
\begin{equation}\label{eq:PS}
    \rm P_k(k) = \left< {\left| \boldsymbol{\delta}_{IC,\boldsymbol{k}}\right|}^2\right>_{|\boldsymbol{k}|=k}~,
\end{equation}
with 
\begin{equation}
    \boldsymbol{\delta}_{\text{IC},\boldsymbol{k}}\propto\ \sum \delta_{\rm IC}(\boldsymbol{x})e^{-i\boldsymbol{k}\cdot \boldsymbol{x}}, \boldsymbol{k}\neq0,
\end{equation}
where the angle brackets denote the average $\langle \rangle$ considering wavevectors $\boldsymbol{k}$ of the same magnitude $\rm k$.
The specific Fourier convention (and so the normalization of Eq. \ref{eq:PS}) is arbitrary in this case, since we only consider dimensionless power-spectrum ratios ($\text{PSR}=\rm P_{\text{k,reconstructed}}/P_{\text{k,true}}$).
The power spectra of the reconstructed ICs can be used to diagnose the scale-dependence of the reconstruction. 
For example, good agreement at low wavenumbers (large scales) indicates that the linear information from the ICs is successfully recovered even with the presence of cosmic variance.

Our second metric, the cross-correlation coefficient, is defined as:
\begin{equation}
    \rm CCC(k) = \frac{\langle\boldsymbol{\delta}_{IC,\boldsymbol{k},true}\cdot \boldsymbol{\delta}_{IC,\boldsymbol{k},reconstructed}^* \rangle _{|\boldsymbol{k}|=\text{k}}}{\sqrt{\rm P_{k,true}\cdot P_{k,reconstructed} }}
    \label{eq:CCC}
\end{equation}
where the average $\langle \rangle$ is again between wavevectors $\boldsymbol{k}$ of the same magnitude $\rm k$.
The CCC is bounded in $[-1,1]$ and provides a normalized metric for evaluating phase differences in two fields. A CCC of 0 indicates no correlation, while a CCC of 1 (-1) indicates perfect correlation (anti-correlation). Throughout the text we define k$_{0X}$ = CCC$^{-1}$(X/10). For example, k$_{05}$ (k$_{08}$) corresponds to the wavemode at which the CCC between the true and reconstructed ICs is 0.5 (0.8).  We take k$_{08}$ as a (somewhat arbitrary) threshold for a "good" reconstruction (c.f. \citealp{Doeser+2024}).\footnote{We performed several additional tests for the evaluation of the recovery, including testing the Gaussianity of the samples and  posterior predictive tests. These tests can be found in Appendix \ref{appndx:sec:statistical_tests}. }  % 

We first quantify the performance of our IC reconstruction using each tracer field individually, under different observational assumptions.  We then show how the recovery improves when using both fields together, assuming fiducial observational parameters.

\subsection{IC recovery from only galaxy maps}

\begin{table*}[htb!]
\centering
\caption{Equivalent selection thresholds for galaxy tracers at $z=8$.}
\label{table:uv_lya}
\begin{tabular}{c c c c c c}
\hline\hline
\noalign{\smallskip}
 $M_{\rm UV}$ & $m_{\rm UV}$ & $L_{\rm Ly\alpha} \, \rm [erg/s]$ & $F_{\rm Ly\alpha} \, \rm [erg/s/cm^2]$ & $\tilde{L}_{\rm Ly\alpha} \, \rm [erg/s]$ & $\tilde{F}_{\rm Ly\alpha} \, \rm [erg/s/cm^2]$   \\
 
\hline
\noalign{\smallskip}
-17  & 32.6 & $2.74\times 10^{41}$ & $3.42 \times 10^{-19}$ & $1.72 \times 10^{42}$ & $2.13 \times 10^{-18}$  \\

   -18 & 31.6 & $5.25\times 10^{41}$ &$6.50 \times 10^{-19}$ &  $3.13 \times 10^{42}$  & $3.87 \times 10^{-18}$  \\
   
   -19  & 30.6 & $1.00\times 10^{42}$ & $1.24 \times 10^{-18}$& $5.58 \times 10^{42}$ & $6.91 \times 10^{-18}$ \\
   -20 & 29.6 &$1.91\times 10^{42}$ &$2.36 \times 10^{-18}$  &  $1.02 \times 10^{43}$ & $1.26 \times 10^{-17}$  \\
\hline
\end{tabular}
\tablefoot{
    Given selected absolute magnitudes $M_{\rm UV}$, we show the corresponding AB apparent magnitude $m_{\rm UV}$, and mean $\rm Ly\alpha$ luminosities and fluxes ($L_{\rm Ly\alpha}$ and  $F_{\rm Ly\alpha}$) using the model of \cite{Gagnon+2026}. We also report the $\rm Ly\alpha$ luminosity and flux limits ($\tilde{L}_{\rm Ly\alpha}$ and  $\tilde{F}_{\rm Ly\alpha}$) that are required to reach a similar number density of a corresponding $M_{\rm UV}$-selected galaxy (obtained via abundance matching). Line-of-sight IGM absorption features are omitted for this baseline comparison.}
\end{table*}

\begin{figure*}[htb!]
    \centering
    \includegraphics[width=0.66\columnwidth ]{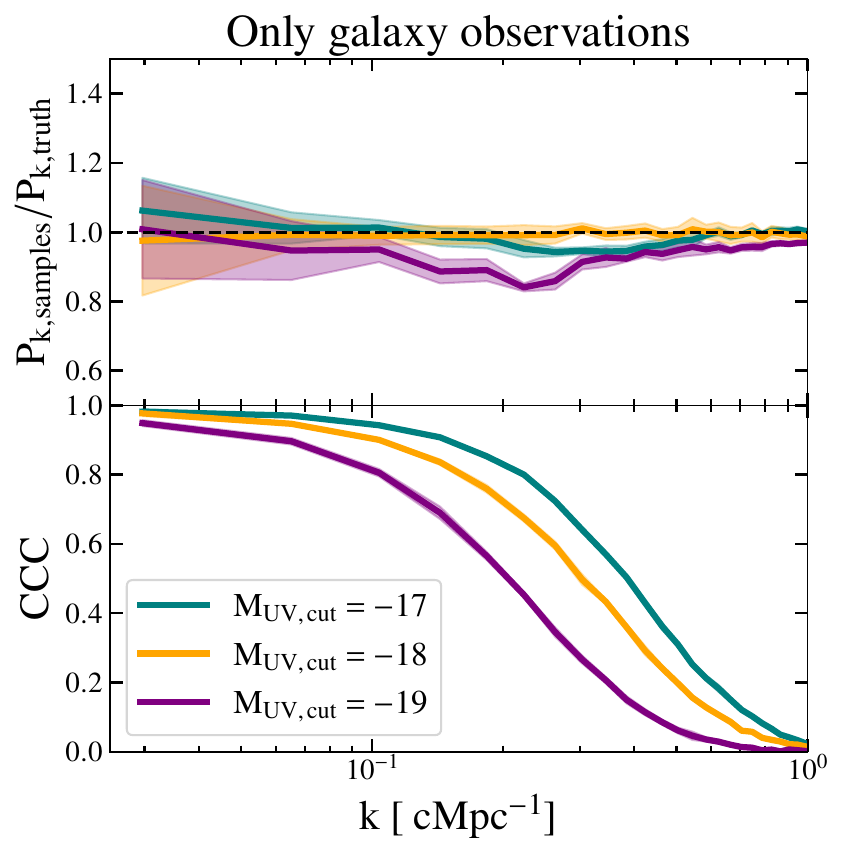}
    \includegraphics[width=0.66\columnwidth]{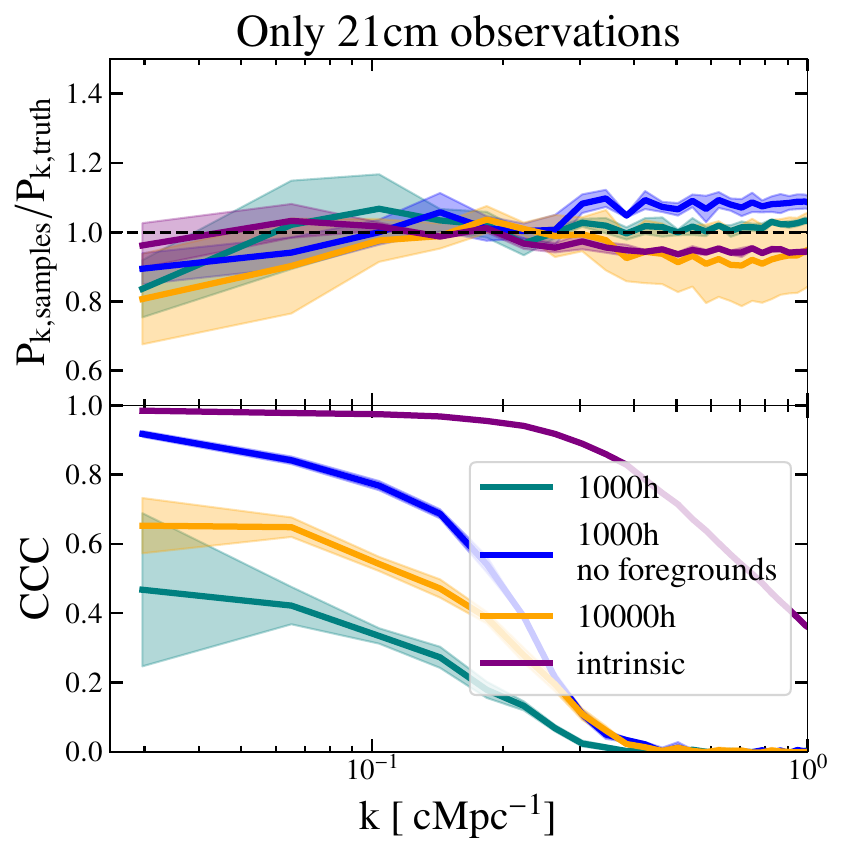}
    \includegraphics[width=0.66\columnwidth]{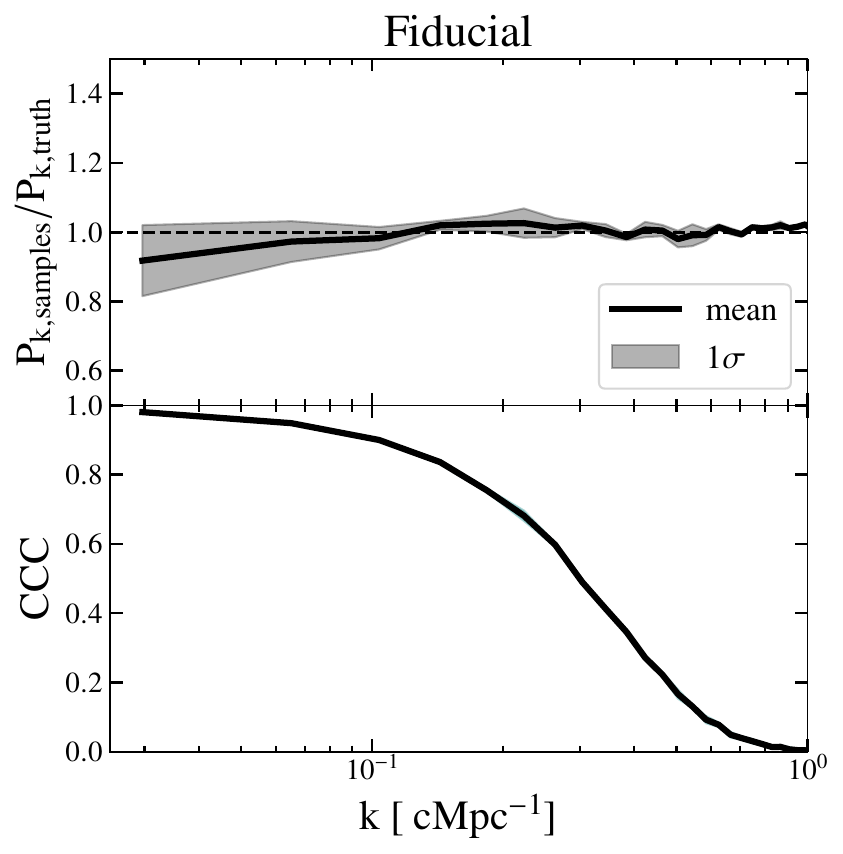}
    
    \caption{The IC power spectrum ratios (PSRs) between samples from {\tt TILING} and the ground truth are shown in the \textit{upper panels}, while the corresponding cross-correlation coefficients (CCCs) are shown in the \textit{lower panels}.  These are computed from 10 test-set realizations, applying {\tt TILING} to obtain 10 posterior samples for each test-set realization, and averaging the results.  Solid lines correspond to the mean while the shaded regions correspond to the root mean square scatter around the mean.
    {\it Left panel}: results from only using galaxies as observational tracers for three different $M_{\rm UV,cuts}\in\{-19,-18,-17\}$ (purple, orange and teal lines respectively). The PSRs are within 10\% for most of the $k$-range, though the shallowest galaxy map has errors up to $\sim$20\%. The CCCs improve with increasing survey depth, with $\rm k_{08,gal}=\{0.11, 0.17, 0.22\}\, cMpc^{-1}$ respectively. {\it Middle panel}: results from only using the 21cm signal as an observational tracer. We show the accuracy of the reconstruction considering the intrinsic signal, the signal including noise from 1000h of observations with SKA-low AA* (no foregrounds), and for the full observational pipeline including foreground-wedge excision for 1000h and 10000h noise, as the purple, blue, teal and orange curves, respectively. While the intrinsic information of the 21cm signal about the ICs is significantly better than for any of the galaxy maps, after sampling with realistic antenna configurations, small-scale information is lost. {\it Right panel}: Results from our fiducial multi-tracer observation that consists of the combination of $M_{\rm UV,cut}=-18$ galaxies and SKA AA* 1000h noise + foreground wedge excision. The result is very similar to the galaxy-only case, although adding the 21cm signal reduces the cosmic variance PSR error on large scales.  This highlights the need for complementary observations when interpreting 21cm tomography in this context.
    \label{fig:psccc_solo_and_joint}
    }
\end{figure*}

\begin{figure*}
    \centering \includegraphics[width=0.92\linewidth]{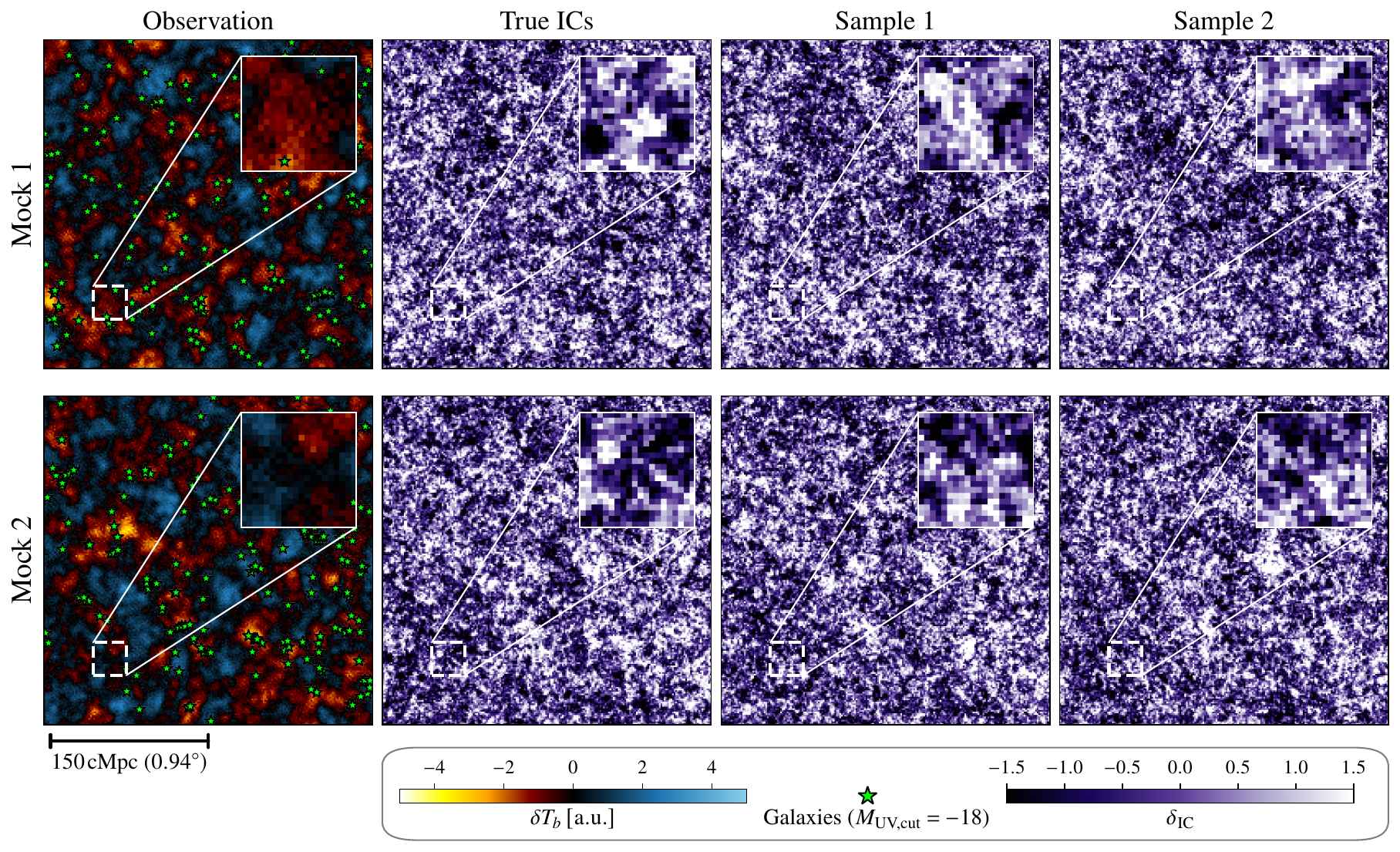}
    \caption{2D slices (width 1.5 cMpc) through the relevant fields, showing example reconstructions for two test-set realizations (first and second row) for our fiducial observational scenario. 
    \textit{First column}: the mock observed galaxies ($M_\text{UV,cut}=-18$) are shown as green stars, together with the mock observed 21cm signal from a 1000h SKA AA* observation including foreground contamination (as in Fig. \ref{fig: 21cm_obs_effects}, we subtract the mean ($4.69 \, \rm mK$) and divide by the rms (286 mK) to standardize the brightness temperature field). 
    As noted before the observed galaxies preferentially trace the large-scale ionized IGM patches where the mean-removed 21cm signal is negative.
    \textit{Second column}:
    the corresponding true ICs for each realization.
    \textit{Third} and \textit{fourth columns}:
    two samples from \texttt{TILING}. As expected from the PSRs and the CCCs of the right column of Fig. \ref{fig:psccc_solo_and_joint}, both samples show very good agreement with the corresponding ground truths in the second column on moderate and large scales.  The zoom-ins highlight small-scale differences.  
    n.b. all IC fields are divided by 2.9 to visually match the range shown in Fig. \ref{fig:main_visual_smoothed}.}\label{fig:main_visual}
% \end{figure*}

% \begin{figure*}
    \centering \includegraphics[width=0.92\linewidth]{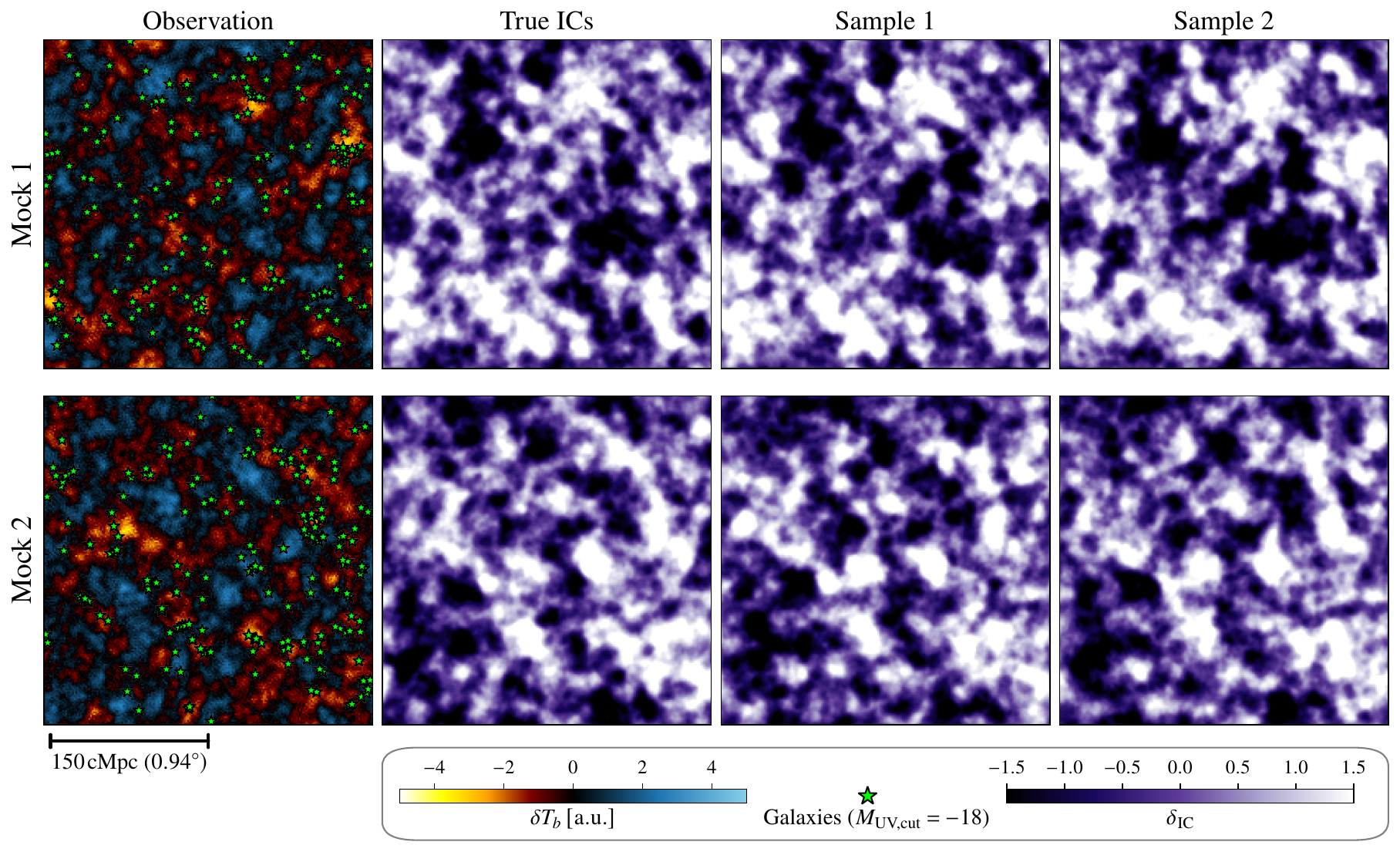}
    \caption{Same as Fig. \ref{fig:main_visual}, but applying a low-pass filter of width $\text{k}=0.2\, \rm cMpc^{-1}$ to the ICs  in order to visually highlight the recovery accuracy on medium and large scales. % with a second-order Butterworth filter. 
See also \url{https://raw.githubusercontent.com/nikos-triantafyllou/nikos-triantafyllou.github.io/refs/heads/main/images/TILING_samples.gif} for more {\tt TILING} samples conditioned on mock 1. }
    \label{fig:main_visual_smoothed}
\end{figure*}

\begin{figure*}[htb!]
    \centering
    \includegraphics[width=0.66\columnwidth]{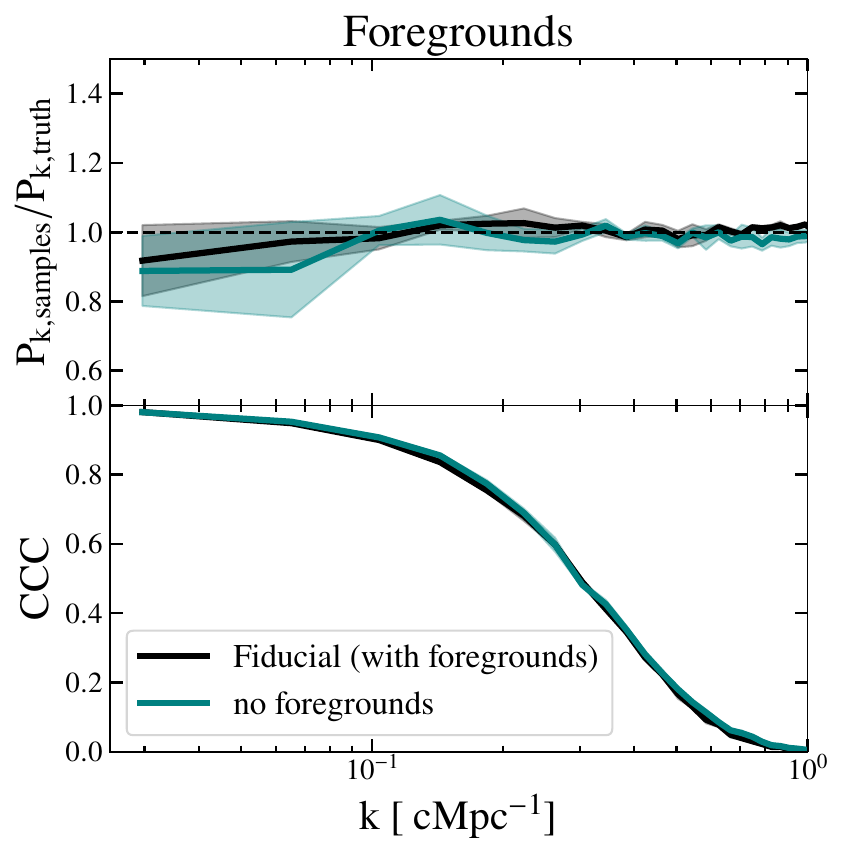}
    \includegraphics[width=0.66\columnwidth]{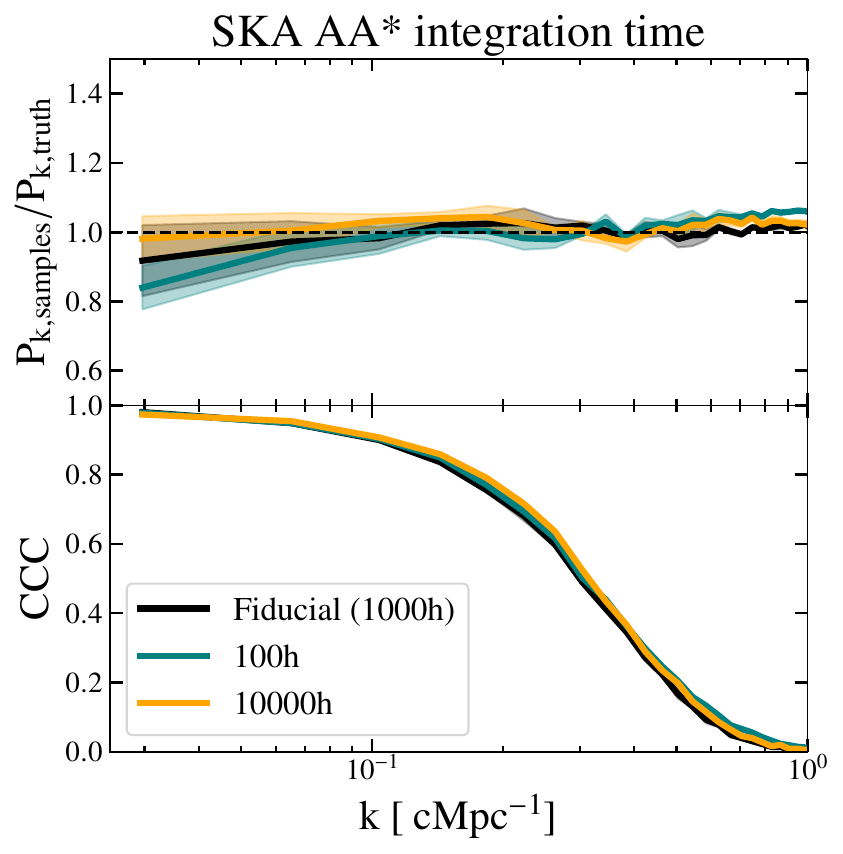}
    \includegraphics[width=0.66\columnwidth]{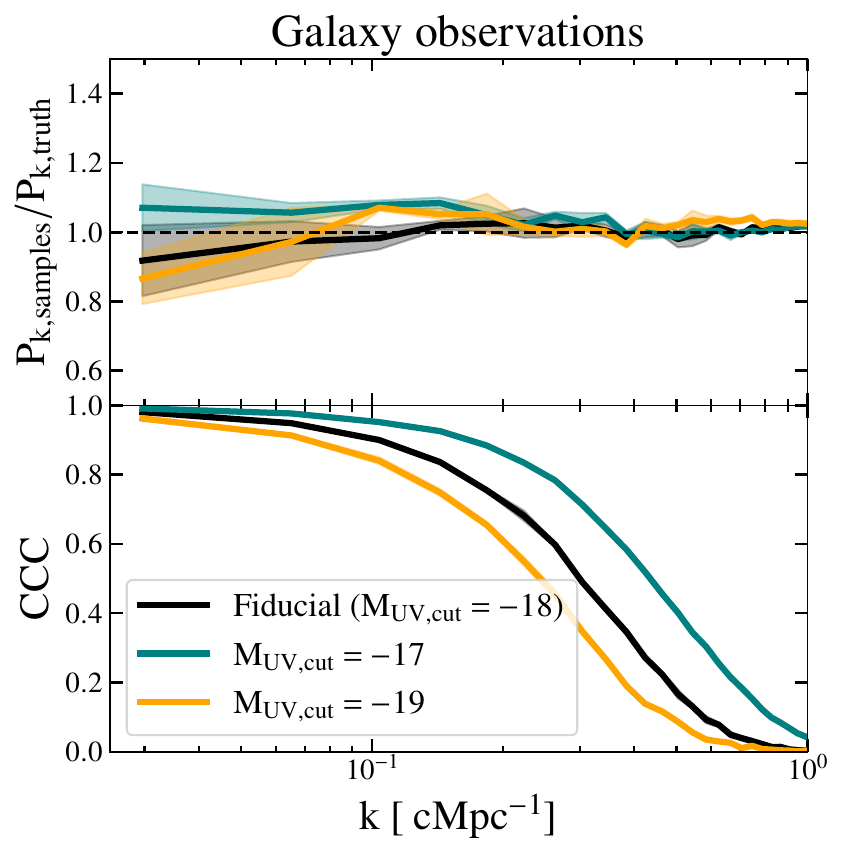}
    \hfill
    \caption{Same as Fig. \ref{fig:psccc_solo_and_joint} but for variations of the observational strategies around the fiducial (shown in black in all panels). {\it Left column}: results considering the fiducial observational scenario with and without foreground-wedge excision. {\it Middle column}:  results considering different SKA AA* integration times: 100h, 1000h, 10000h, while keeping the fiducial $M_{\rm UV,cut}=-18$ depth of the galaxy field. {\it Right column}: results considering the fiducial SKA AA* noise and foregrounds but varying the galaxy survey depth, including $M_{\rm UV,cut}=-17$ and $M_{\rm UV,cut}=-19$ (teal and orange lines respectively).  As discussed in the text, most of the IC recovery power in our fiducial multi-tracer set-up comes from the galaxy maps, though 21cm helps to constrain the recovered power spectra, especially for shallow galaxy surveys.}
    \label{fig:psccc_fid_var}
\end{figure*}

To explore the performance of the pipeline under varying observational thresholds, we consider three representative UV magnitude cuts: $M_{\text{UV,\,cut}} = \{-17, -18, -19\}$. These cuts correspond to mock galaxy number densities of $\sim$ $4.4 \times 10^{-3}$, $1.7 \times10^{-3}$, and $4.6\times 10^{-4}$ $\text{cMpc}^{-3}$, respectively.  In practice, such selection functions should be tailored to match specific observational surveys.  For example, line-intensity surveys would have a corresponding cut on the line luminosities, while spectroscopic follow-up would involve sampling from the corresponding observational selection function.  As this work is a proof-of-concept, we do not focus on any existing or planned survey.   We note only that the above magnitude cuts roughly span the range achievable with deep imaging with {\it JWST} (e.g., \citealp{Bagley2024, Eisenstein+2026, Finkelstein+2025, Donnan+2024}) and shallower/wider grism or slit-spectroscopic surveys with instruments like the \textit{RST}, VLT/ELT, Subaru PFS and {\it JWST}  (e.g., \citealp{Maiolino+2020, Roman+2025, Bagley+2026, Wold+2024, Perez+2023}).  The latter typically rely on Lyman alpha for source identification and so for reference we show the mean Lyman alpha luminosities and fluxes corresponding to our UV magnitude cuts in Table \ref{table:uv_lya}, computed using the stochastic relations in \cite{Gagnon+2026}.  In future work we will quantify the dependence of the reconstruction on observational selections and survey geometries.  We note only that our observational selections are comparable to, or more conservative than, similar works in the literature \citep[e.g.,][]{Maitra+2026}.

In Fig. \ref{fig:psccc_solo_and_joint} (left panel) we show the PSRs alongside the CCCs between the reconstructed and ground-truth power spectra, derived from 10 test-set realizations and 10 posterior samples per realization.  Solid colored lines correspond to the three different UV cuts as indicated in the caption. In the top panel, the recovered power spectra demonstrate strong agreement with the ground truth, with the mean and $1\sigma$ credible intervals remaining within $10\%$ for most of the range.  %
In the bottom panel, as expected, the CCC increases systematically with survey depth, yielding $\rm k_{08}\in\{0.11, 0.17,0.22\} \, cMpc^{-1}$ for $M_{\rm UV,cut}\in \{-19,-18,-17\}$.  We note that the improvement from $M_{\rm UV, cut}$ = -18 $\rightarrow$ -17 is much more modest than the improvement from $M_{\rm UV, cut}$ = -19 $\rightarrow$ -18.  Below we take $M_{\rm UV, cut}$ = -18 as the fiducial galaxy survey choice.

\subsection{IC recovery from only 21cm tomography}

Following the same format, in the middle panel of Fig. \ref{fig:psccc_solo_and_joint} we show results for different choices of  integration time with SKA-low AA*: $\sim$1000h ({\it teal curves}) and $\sim$10000h ({\it orange curves}) corresponding to 167 and 1667 days of observation respectively. The former choice is typical of a "fiducial" deep field, while the latter would be typical of a multi-year tomographic survey (e.g., \citealp{CD_SWG+2026}).  For reference, we also show the reconstruction accuracy for a 1000h observation free of foreground contamination ({\it blue curves}), as well as one inferred directly from the intrinsic 21cm signal (i.e., without the observational effects discussed in Sect. \ref{sect:21cm_obs}; {\it purple curves}).

Looking at the top panel we see that all observational scenarios reasonably recover the correct amplitudes of the IC power spectra (to within tens of percent).  However, the CCCs in the bottom panel show that the observational scenarios strongly impact the recovered {\it phases}.  Recovery using the intrinsic signal ({\it purple curves}) works extremely well, achieving $\rm k_{08}\approx0.4\, cMpc^{-1}$ and $\text{CCC} \approx 1.0$ across large and intermediate scales ($\rm k\leq 0.3 \, cMpc^{-1}$).  However, the reconstruction quality decreases significantly when we account for $uv$ sampling of the interferometric array.  Even under the optimistic assumption of being able to clean foregrounds entirely inside the wedge ({\it blue curves}), the CCC only rises above 0.8 on large scales, $\rm k_{08}\approx0.09\, cMpc^{-1}$.  When the wedge is excised, the CCC never goes above 0.8, although increasing the integration time does relatively improve the recovery, with $\rm k_{04}$ increasing from $0.07$ to $0.18\, \rm cMpc^{-1}$. 

The results in this section indicate that the $uv$ sampling of SKA-low is the largest impediment to accurate phase reconstruction, as small scales are not sampled with a sufficient number of long baselines.  The telescope layout results in a characteristic cutoff when the $\text{CCC} \to 0$ at $\text{k} \approx0.35$. We note that using the AA4 configuration yields similar results (c.f. Appendix \ref{sect:ska_obs}); likely not appreciably changed even with the proposed deferral of 50 stations  (c.f. \citealt{Breitman+2026} for a related discussion). This highlights the need for complementary observations, such as galaxies, when interpreting 21cm tomographic observations.

\begin{figure*}
    \centering \includegraphics[width=1\linewidth]{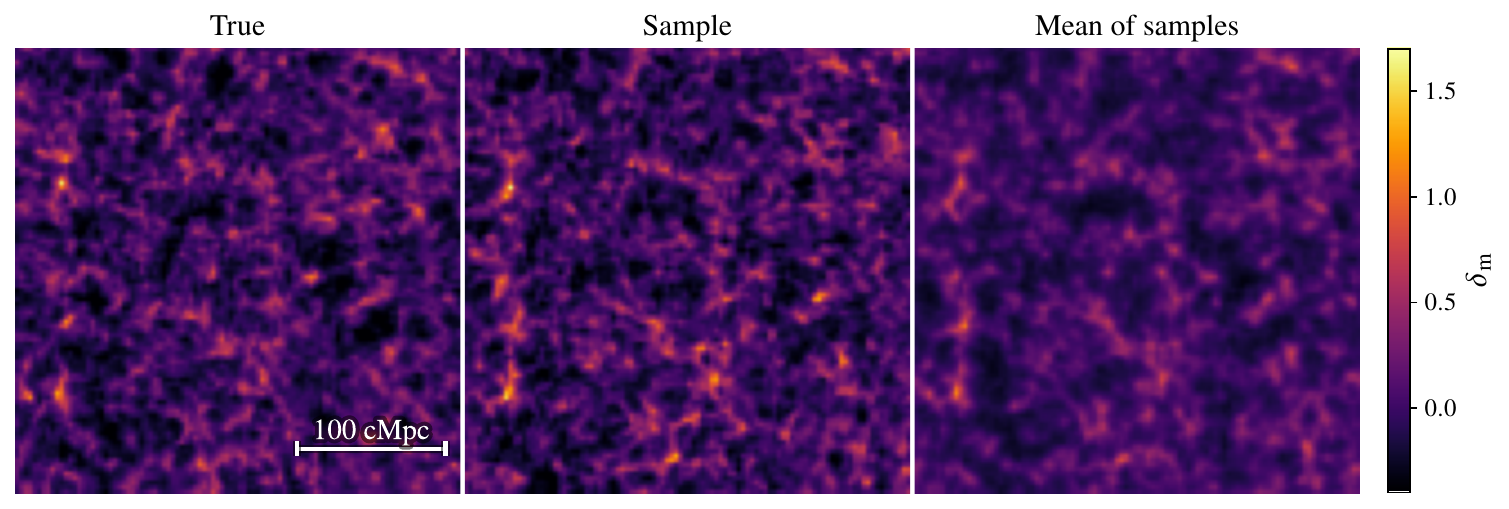}
    \includegraphics[width=1\linewidth]{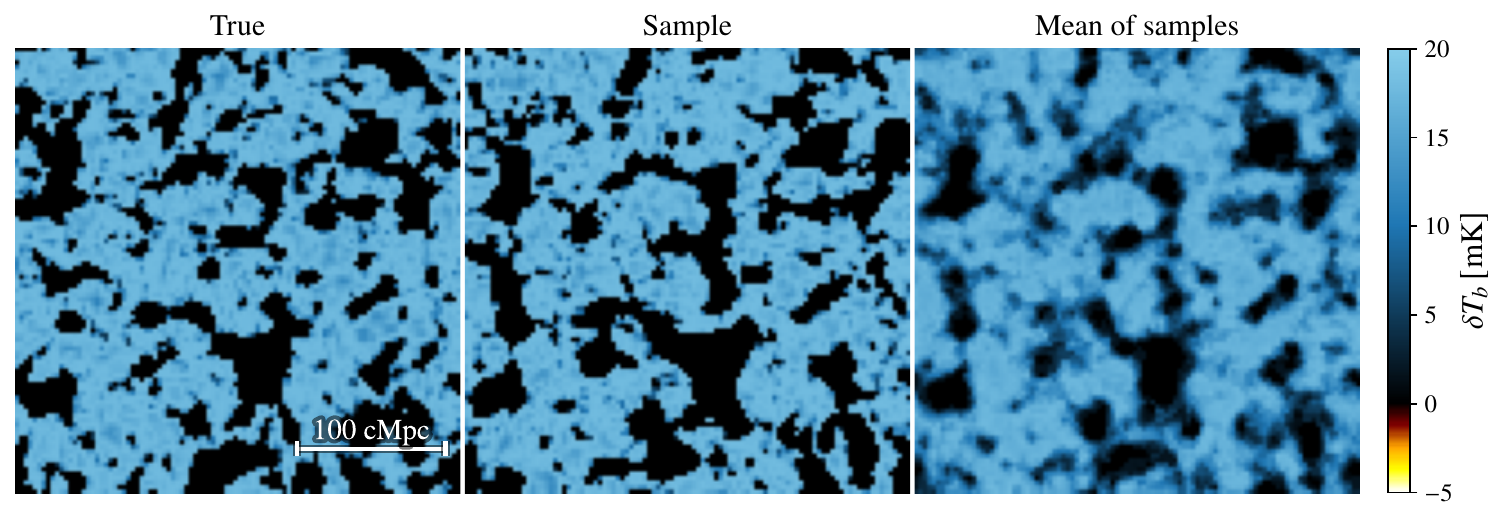}
    \caption{Posterior re-simulations of constrained ICs for a test-set realization. {\it Top row}: Evolved density field (averaged over $9\,\rm cMpc$ along the LoS). {\it Bottom row}: 21cm signal (averaged over $6\,\rm cMpc$ along the LoS). From left to right, the columns display the true fields, an individual posterior sample, and the mean of 10 posterior samples. Individual samples closely match the ground truth while naturally preserving correct statistical properties. The mean is smoother on the small scales where the uncertainty is highest.}
    \label{fig:evolved_fields}
\end{figure*}

\subsection{Multi-tracer synergy: joint inference}
In the previous sections, we presented the summary statistics results for two observational tracers independently. Here we perform a joint multi-tracer inference by combining both datasets under our fiducial observational scenario: a $M_{\rm UV, cut}=-18$ galaxy catalog and a SKA AA* observation of 1000h affected by foreground wedge excision.

In the right panel of Fig. \ref{fig:psccc_solo_and_joint} we show our two IC recovery metrics corresponding to this joint inference. Both the PSRs and the CCCs are very similar to the corresponding galaxy-only result, although adding the 21cm signal helps reduce the cosmic variance error on large scales.

For this fiducial case we show the recovery visually in Fig.  \ref{fig:main_visual}.  
The two rows correspond to two different realizations from the test set.  In the first column, we show the mock galaxy + 21cm  observations.  In the second column we show the corresponding true ICs that were used to generate the mock observations.  In the final two columns, we show the IC recovery from two calls to {\tt TILING}, conditioned on the mocks shown in the first column.  All fields are 2D slices of width $1.5\, \rm cMpc$ through the corresponding 3D fields that are used in {\tt TILING}.

From the first column we see again how the bright, observable galaxies preferentially trace the ionized regions where the brightness temperature is negative after mean subtraction.  These are hosted by large-scale matter overdensities seen in the second column.  Comparing the last two columns to the ground truth shown in the second column, we see that the recovery works very well on large scales though the zoom-in shows small-scale discrepancies (as could be inferred from the CCC shown in the bottom-right panel of Fig. \ref{fig:psccc_solo_and_joint}).

To highlight the quality of the recovery on large scales, in Fig. \ref{fig:main_visual_smoothed} we show the same IC fields as in Fig. \ref{fig:main_visual}, but smoothed  
on $\text{k}=0.2\,\text{cMpc}^{-1}$ scales using a second-order Butterworth filter (\citealp{butterworth+1930}):
\begin{equation}
\delta_{\text{smoothed}}(\boldsymbol{k}) = \delta_{\text{IC},\boldsymbol{k}} \cdot \frac{1}{\sqrt{1 + (|\boldsymbol{k}| / 0.2)^{4}}}.
\end{equation}
The samples from {\tt TILING} are nearly identical to the ground truth on these scales.

The reader is referred to Appendix \ref{sec:appndx:tests_gauss} for additional tests of recovery in the fiducial case,  where we show the samples' consistency with the assumed Gaussianity. Furthermore, in Appendix \ref{appndx:sec:coverage} we perform pixel-based coverage tests for 100 test-set realizations, showing a well-calibrated posterior. 

Finally, in Fig. \ref{fig:psccc_fid_var} we quantify how our multi-tracer recovery depends on observational effects. In the left column, we show results with and without 21cm foreground wedge removal (i.e., ignoring foreground contamination).  We see that the recovery is largely unaffected by wedge excision, due to the fact that in our fiducial set-up, the galaxy maps dominate the constraining power.

In the middle column of Fig. \ref{fig:psccc_fid_var} we vary the integration time with SKA-low, showing results for 100h, 1000h, and 10000h.  The CCCs are largely unaffected, though the power spectrum recovery does degrade by up to 10\% for 100h.  This points to the conclusion that although the galaxy maps determine the phases of the recovered ICs in our fiducial multi-tracer set-up, the 21cm maps help to constrain the Fourier amplitudes (i.e., the power spectrum).

This is further evidenced in the right column, where we vary the galaxy UV magnitude threshold.  Comparing these curves with the analogous ones in the left column of Fig. \ref{fig:psccc_solo_and_joint}, we see that the CCC and in particular the $\rm k_{08}$ values are unchanged.  However, the inclusion of 21cm improves the power spectrum recovery, especially for the shallowest survey with $M_{\rm UV,cut}=-19$.  The galaxy-only recovery resulted in mean PSR errors of up to 20\%, even at moderate scales of $\text{k}\sim0.2$ cMpc$^{-1}$.  Adding 21cm maps to the recovery pipeline decreases this error to a few percent.

% SECT 5: RESULTS PART 2 ================================================
% ===============================================================
% ===============================================================

\section{Evolved fields and wedge reconstruction}\label{sec:evolved_fields}

One powerful application of {\tt TILING} is to use the inferred posterior samples as initial conditions for cosmological simulations.  By ``re-simulating'' the observed volume, one is able to study its full history, uncorrupted by observational effects.  In this section we briefly show one such application: recovering the 21cm foreground wedge.\footnote{In a companion paper, \ct{} we go into detail about using {\tt TILING} to recover galaxy and ionization field evolution of observed volumes.}

We use the IC samples produced by {\tt TILING}, as described in the previous section, as input to {\tt 21cmFAST} in order to re-simulate the corresponding late-time constrained realizations of the density and 21cm fields. We use slightly different parameter choices  for this re-simulation.  Specifically, these include: (i) a sharp-$k$ filter instead of the default exponential spherical tophat filter from \cite{Davies+2022} for ionizing emissivity fields; (ii) ignoring sub-grid recombinations; (iii) ignoring the contribution of molecularly-cooled galaxies; (iv) assuming a saturated spin temperature ($T_s \gg T_{\rm cmb}$); (v) fixing
$M_\text{turn, low}=10^5\rightarrow 10^{8.7}$ M$_\odot$;  (vi) using a grid resolution of 
$\rm 3\, cMpc$; and (vii) computing radiation fields using the mean conditional halo abundances rather than the stochastic, discrete source model introduced in v4.

These different modeling choices further test the robustness of
${\tt TILING}$, as the output simulator is effectively different than what was used in training. Accurate IC recovery therefore would require {\tt TILING} to be able to generalize to a different simulation (albeit one done using the same code).   Fully quantifying the impact of model misspecification would require testing results against multiple different simulation codes during inference (e.g., \citealp{Jo+2025, Meriot+2026}), and is beyond the scope of this work.

The first row of Fig. \ref{fig:evolved_fields} shows one example true late-time density field (\textit{left panel}), alongside a re-simulated posterior sample (\textit{middle panel}) and the mean across 10 samples (\textit{right panel}).  The mean appears smooth because small scales are weakly constrained, which effectively translates to noise in each realization that averages out in the mean. The random sample and the mean both recover medium to large scales in both fields very well. We provide the corresponding halo mass function in Appendix \ref{appndx:sec:statistical_tests} as a non-linear quality metric of the recovered density fields.

The second row of Fig. \ref{fig:evolved_fields} shows similar behavior for the constrained realization of the 21cm signal.  We caution again that here we fix the astrophysical parameters, assuming those to be recovered from observations with a comparably smaller uncertainty than the ICs.  We will explore co-varying astrophysical parameters in future work. 

\begin{figure*}
    \centering
    \includegraphics[width=0.99\columnwidth]{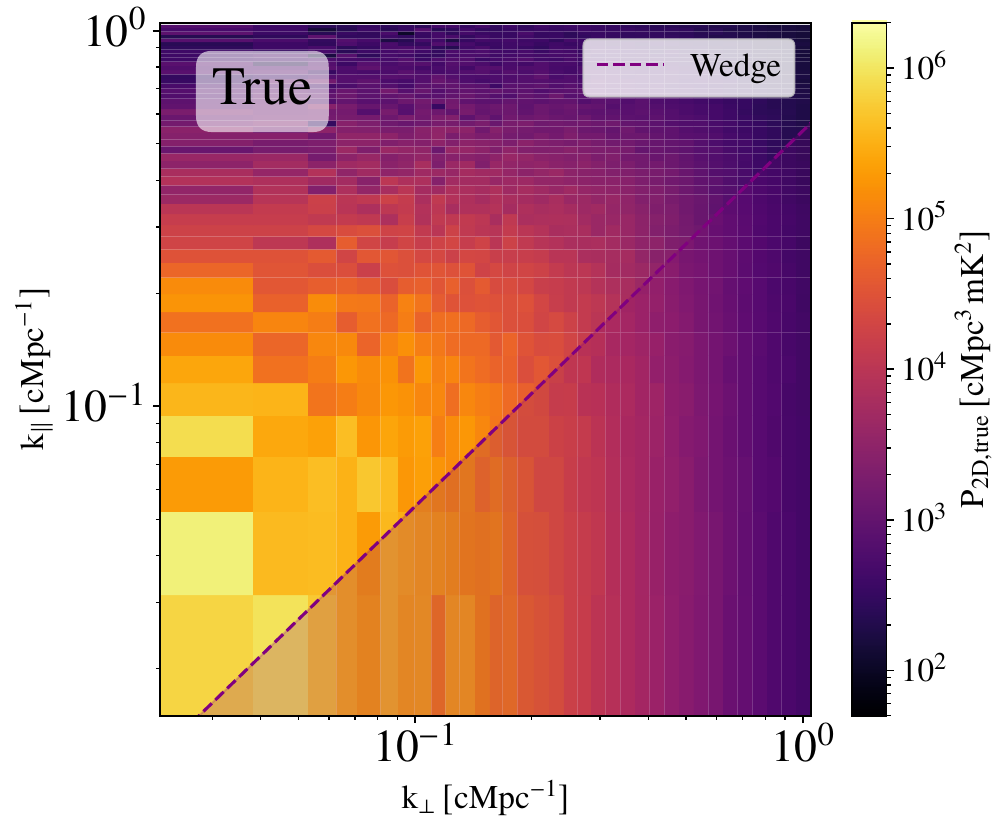}
    \includegraphics[width=0.99\columnwidth]{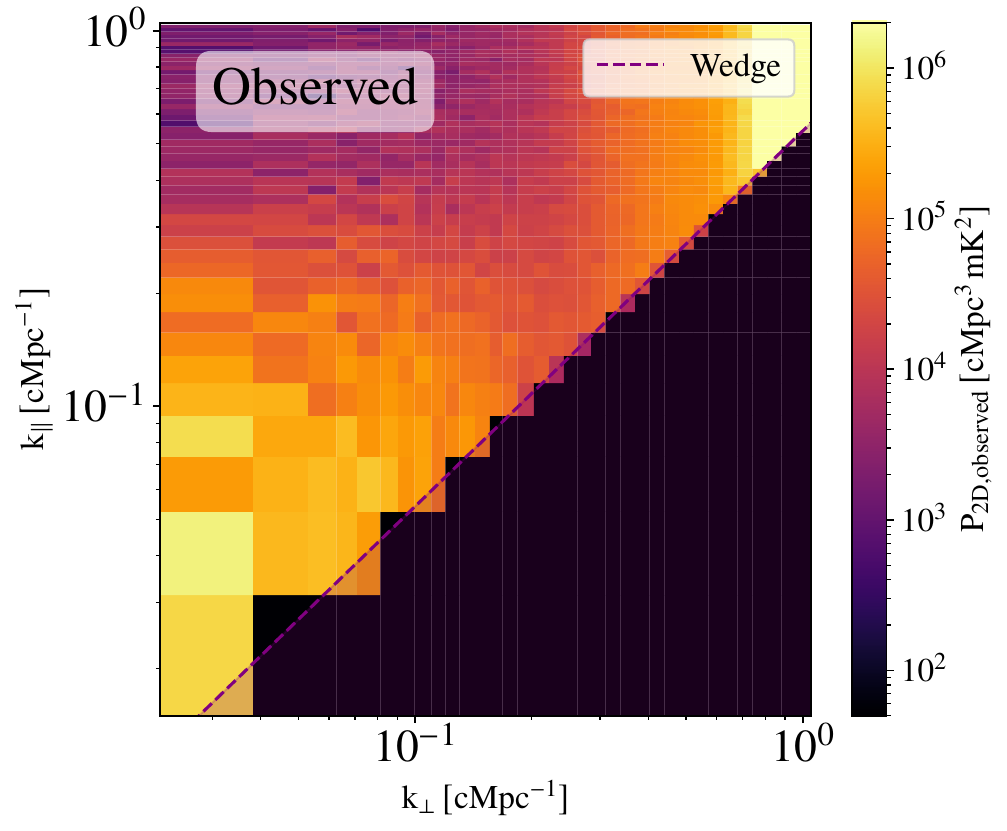}
    \includegraphics[width=0.99\columnwidth]{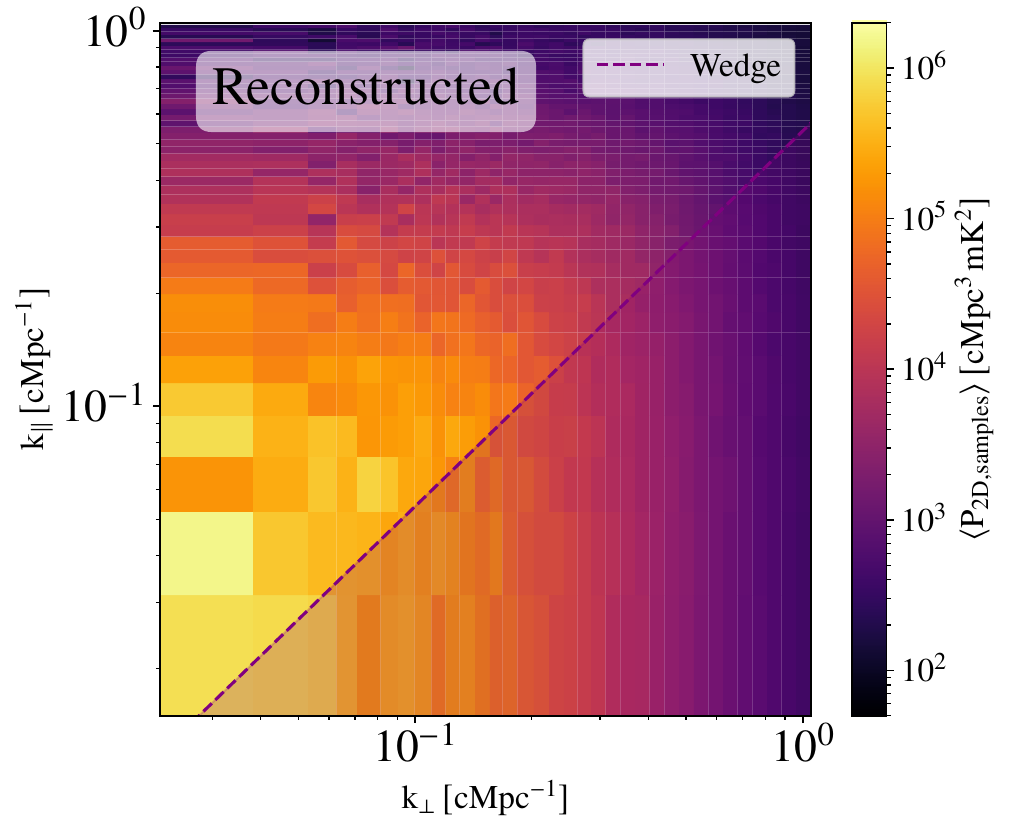}
    \includegraphics[width=0.99\columnwidth]{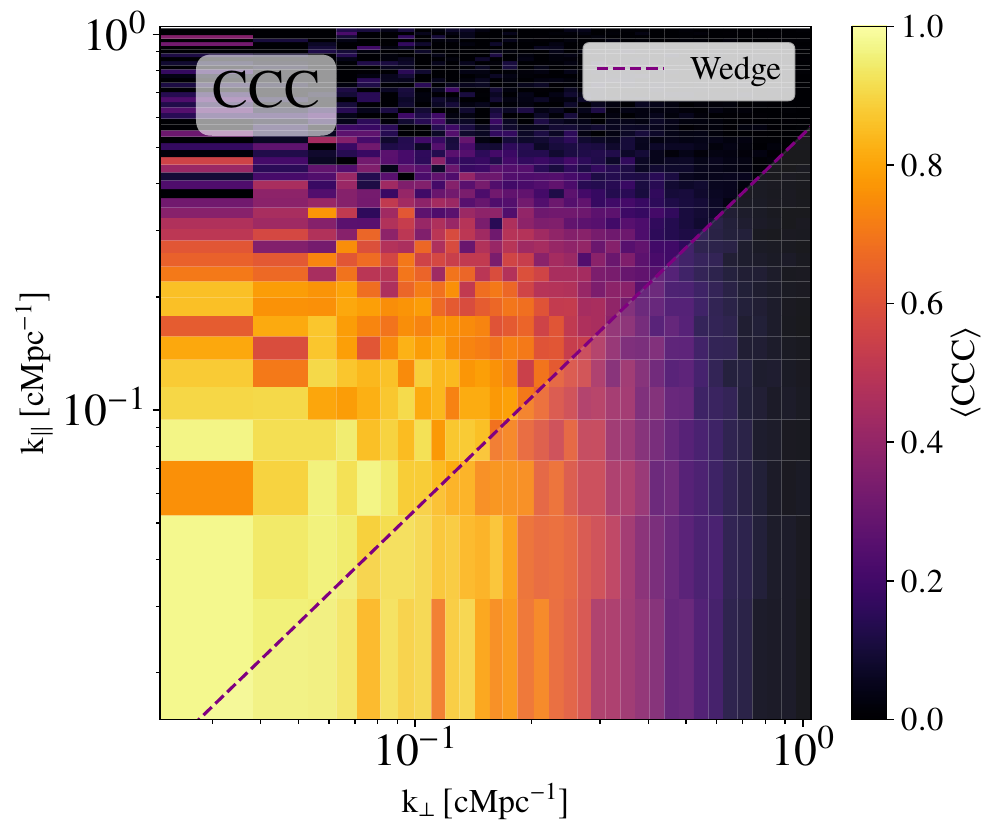}
    \caption{Cylindrical power spectra of 21cm brightness temperature fields at $z=8$ from a test-set realization. \textit{Top panel}: The intrinsic (\textit{left}) and the observed (\textit{right}) signal after SKA AA* 1000h observation and foreground wedge excision. \textit{Bottom panel}: The mean cylindrical power spectra from 10 re-simulated ICs samples (\textit{left}) alongside the mean of their CCCs (\textit{right}). Accurate reconstruction is demonstrated by the similar values of the true and reconstructed power spectra together with their high CCCs ($\text{CCC}>0.5$ for the majority of the wedge). See also Fig. \ref{fig:evolved_fields}. }
    \label{fig:wedge}
\end{figure*}

We quantify the 21cm wedge reconstruction quality in Fig. \ref{fig:wedge}, where we show the 21cm cylindrical power spectra ($\rm P_{2D}$; in $k_\parallel, k_\perp$ space) from one random test-set realization.
In the \textit{top row} we show the PS of the true intrinsic signal (also without RSDs; \textit{left panel}) and the mock observation (\textit{right panel}).
The \textit{bottom left} panel shows the reconstructed $\rm P_{2D}$ as the mean of 10 samples from {\tt TILING}.  Comparing the top and bottom left panels, we see excellent agreement between the true and reconstructed PS, both inside and outside the foreground-dominated wedge.  

We further quantify the quality of PS recovery by plotting the cylindrically-averaged CCCs in the  \textit{bottom right} panel. We see that the wedge (purple shaded region) is recovered with $\text{CCC}>0.5$ accuracy across most of its domain.
These results are roughly comparable to \cite{Chen+2025}, where they used both  approximate models based on EFT as well as a diffusion approach to recover the wedge from foreground-contaminated 21cm mock observations. Our results indicate improved CCCs especially in the $k_\perp$ direction, demonstrating the strength of using galaxies as an additional tracer. Nevertheless, the inherent differences between approaches (discussed previously) do not allow for a clean, apples-to-apples comparison.

\section{Conclusions}\label{sec:conclusions}

The ICs constitute a clean probe for cosmology. 
Reconstructing them recovers the full temporal evolution of an observed volume.
While such reconstruction pipelines are fairly mature at low redshifts, application to the EoR has remained relatively unexplored due to the challenges associated with modeling cosmic radiation fields and a historic dearth of observations.  

In order to prepare for ongoing and future large EoR surveys, here we develop \texttt{TILING}, a hybrid pipeline for field-level, multi-tracer reconstruction of IC fields. \texttt{TILING} pairs a point-estimate NN, %
inspired by \cite{Savchenko+2025}, with a conditional score-based diffusion model that refines this estimate and provides posterior samples of 3D ICs.% 
We train {\tt TILING} on combinations of mock galaxy and 21cm maps at $z\sim8$, with varying observational effects, quantifying the accuracy of recovery.

Our main results are the following:
\begin{itemize}
\item \textbf{High-Fidelity Joint Reconstruction:} \texttt{TILING} can reliably constrain ICs down to small scales ($\rm k \lesssim 0.2\ \text{cMpc}^{-1}$) by performing joint inference from upcoming galaxy and 21cm observations.

\item \textbf{Synergy of galaxy and 21cm maps:} We find that most of the constraining power from {\tt TILING} comes from the galaxy maps, further highlighting the importance for multi-tracer analysis and observations with shared footprints.  This is due to the baseline distribution of the SKA, which is not sensitive to the small scales accessible with galaxy maps.  The 21cm signal as observed with SKA does aid in recovering the IC power spectra.

\item \textbf{Foreground Wedge Recovery:} \texttt{TILING} successfully recovers modes of the 21cm power
spectrum that were excised due to foreground contamination. The foreground modes can be reconstructed with $\text{CCC}>0.5$ across most of their domain, under fiducial observational assumptions.
\end{itemize}

Additionally, our IC reconstruction framework can be used to (i) guide follow-up observations of sub-volumes of interest, (ii) reconstruct the galaxy evolution and corresponding reionization morphology of specific volumes, and (iii) isolate the contribution to reionization by the vast majority of galaxies that will remain unobserved by optical/IR telescopes such as \textit{JWST}. We address (ii) and (iii) in TILING II (\ct{}), where we demonstrate how an IC reconstruction can reduce cosmic variance and predict the joint evolution of reionization and galaxies in specific sub-volumes.

% SECT last NOTES and beautiful sentences to put in the paper =============================================
% ===============================================================
% ===============================================================

\section{Data availability}\label{sec: data avail}
The data underlying this article will be shared on reasonable
request to the corresponding author. Any future publishing of the data will be made available on the corresponding Github page.

% SECT last NOTES and beautiful sentences to put in the paper =============================================
% ===============================================================
% ===============================================================

\begin{acknowledgements}
We gratefully acknowledge computational resources of the Center for High Performance Computing (CHPC) at SNS. We acknowledge CINECA awards under the ISCRA initiative for providing us access to the LEONARDO supercomputer (IsB30\_IC-diff, project no. HP10B1D1F2, PI: Triantafyllou; CNHPC\_1497299, project no. 1497299, PI: Mesinger). This project is supported by Italian Ministerial grant PNRR from National Centre for HPC, Big Data and Quantum Computing CUP E53C22000790001, Spoke 3. This work was recognized as the winning poster at E4 Computer Engineering's "AI, HPC, QUANTUM" event (Milan, September 2025). The authors thank Daniela Breitman for her original contributions to \texttt{tuesday} and her valuable support. We also thank Adrian Liu, Yi Mao and Daniela Breitman for their valuable suggestions that helped improve this work. N.T. expresses his gratitude to David Prelogović for his  insights and guidance during the initial phases of this project.
\end{acknowledgements}

\bibliographystyle{aa}
\bibliography{bib}

% APPENDIX=======================================================
% ===============================================================
% ===============================================================
% \onecolumn
\begin{appendix}

\section{SKA observations}\label{sect:ska_obs}
\begin{figure}[htb!]
    \centering
    \includegraphics[width=1\linewidth]{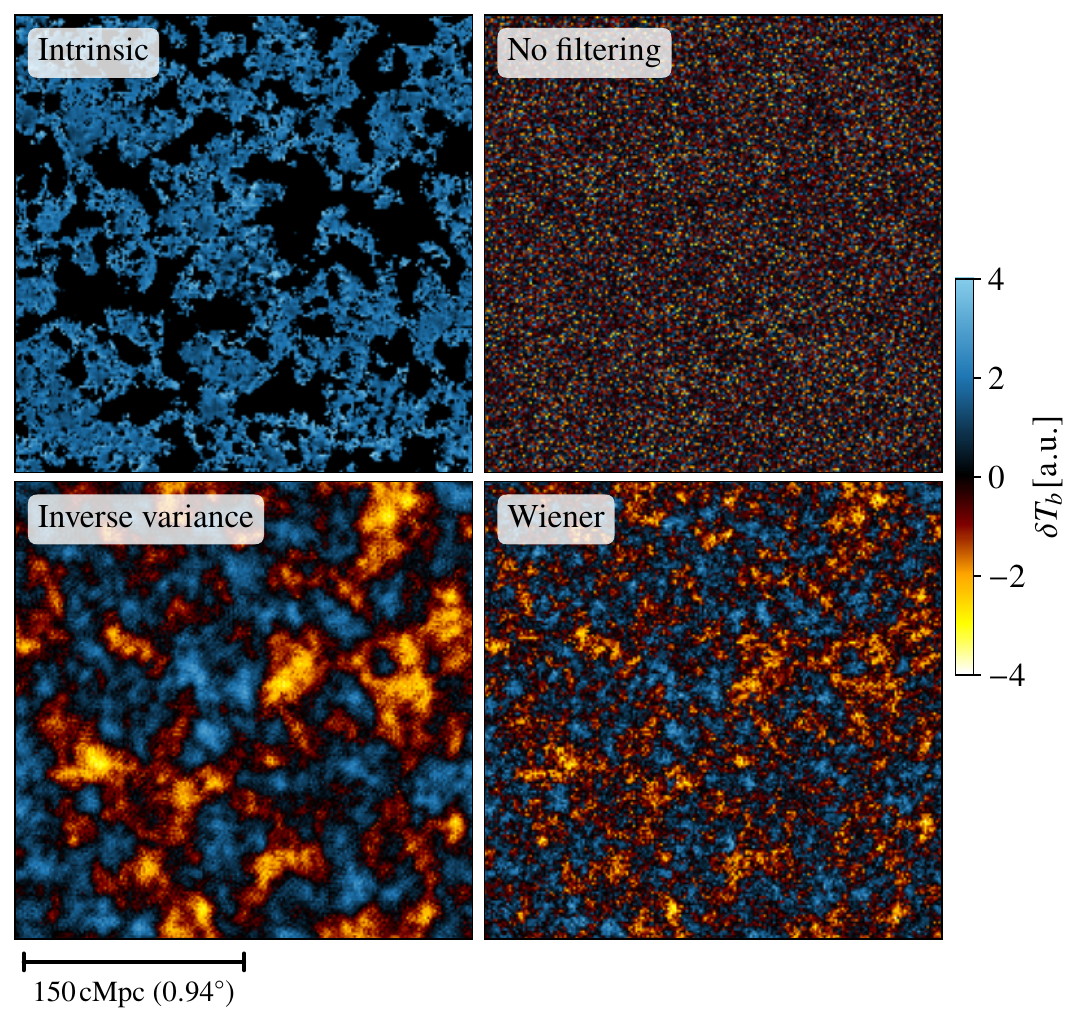}
    \caption{Different filtering schemes for SKA AA* 1000h at $z=8$. Depth of 1.5 cMpc.
    Top panels: The intrinsic signal (left) and the signal with added noise under no filtering scheme (right), showing no apparent structure.
    Bottom panels: The inverse-variance and the Wiener filtered signal (left and right, respectively). With filtering, the morphology of the intrinsic signal is recovered, especially in the inverse-variance case. Despite the visual improvement, the performance of \texttt{TILING} in the PSRs and CCCs was indifferent to the choice of filtering. All fields are divided by their standard deviation for visual comparison ($5.71$, $1220$, $360$, and $348\,\mathrm{mK}$ for the top-left, top-right, bottom-left, and bottom-right panels, respectively). All fields with noise have undergone mean subtraction ($4.69\, \rm mK$) due to the interferometric pipeline.}
    \label{fig:appndx:filtering}
\end{figure}

Here, we show how different filtering schemes in $uv$ space visually affect the SKA 21cm observation.
The top left panel of Fig. \ref{fig:appndx:filtering} shows an example intrinsic\footnote{Note that in this appendix we refer to ``intrinsic'' as the RSDs-free signal, where the noise is added for this comparison.} signal from our dataset, using only one slice of the simulation at $z=8$ corresponding to $1.5\rm\, cMpc$ depth in the LoS. 
After adding noise to the signal in $uv$ space (as described in Sect. \ref{sect:21cm_obs}), transforming directly back to real space without any additional processing results in a highly degraded signal, as illustrated in the second panel of Fig. \ref{fig:appndx:filtering}. This effect could compromise the performance of the NN as features may become difficult to extract. 

\begin{figure}[htb!]
    % \centering
    \hspace{0.8cm}\includegraphics[width=0.9\columnwidth]{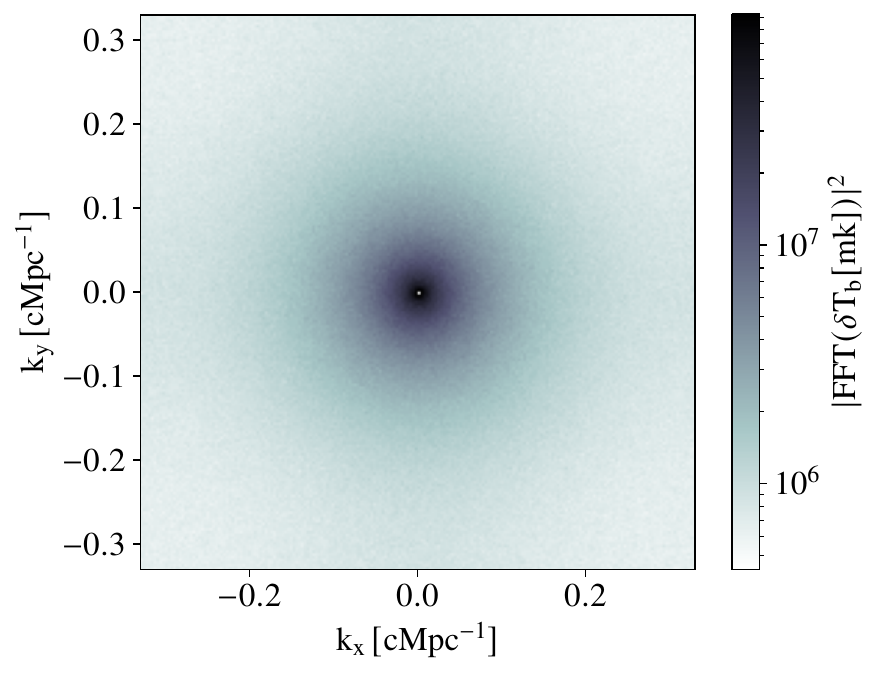}

    \hfill
    \caption{Matrix used for Wiener filtering. Calculated as the median of Fourier transformed 21cm signal realizations at $z=8$ for a range of astrophysical parameters specified in the prior of \ct{}.}
    \label{fig:appendx:Wiener_fixed}
\end{figure}

\begin{figure}[htb!]
    \centering
    \includegraphics[width=0.99\linewidth]{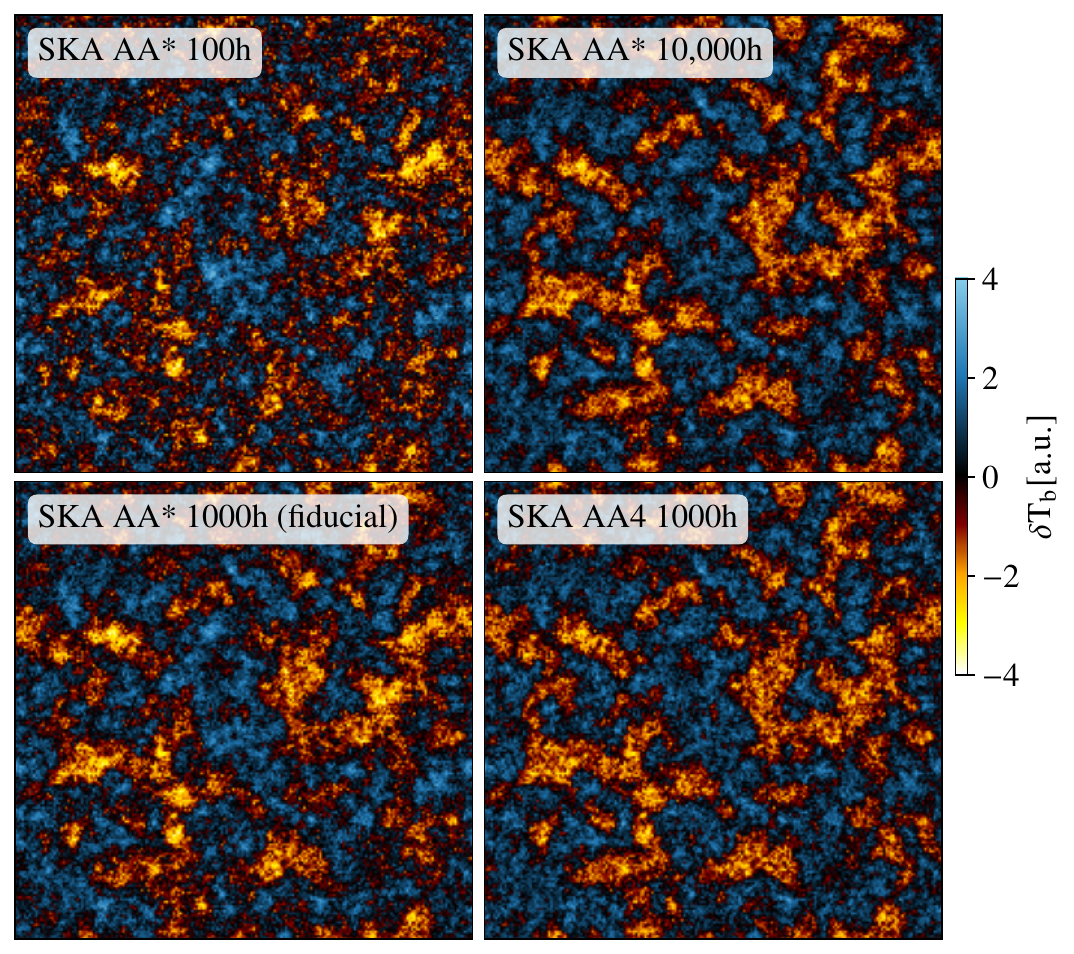}
    \caption{Different Wiener-filtered noise levels for SKA observations at $z=8$, averaged over $7.5\, \rm cMpc $ in the LoS.
        Top panels: perturbations around the fiducial noise level, showing total integration times of 100h (left) and 10,000h (right). 
        A severe reduction in visible structure is evident at 100h, whereas the 10,000h case yields minor visual differences compared to the baseline. 
        Bottom-left panel: The fiducial 1000h observation. Bottom-right panel: The mock signal simulated using the SKA AA4 configuration, showing no significant enhancement in morphological clarity over the fiducial. All fields are divided by their standard deviation for visual comparison ($965$,  $141$, $360$, and $208\,\mathrm{mK}$ for the top-left, top-right, bottom-left, and bottom-right panels, respectively). All fields have undergone mean subtraction ($4.69\, \rm mK$) due to the interferometric pipeline.}
    \label{fig:appndx:noise_levels}
\end{figure}

For this reason, we attempted to weight the $uv$-space modes to reveal structure in the following way for each $uv$-space cell at fixed frequency channel 
$i$:
\begin{equation}
    \delta T^{i}_{b,uv,\text{filtered}}(u,v) = \delta T^{i}_{b,uv}(u,v),\frac{w^{i}(u,v)}{\sum_{u',v'} w^{i}(u',v')}
\end{equation}
where $ \delta {T^i_{b,uv}}$ is the unfiltered 2D signal in $uv$-space, $\delta T^i_{b,uv, \text{filtered}} $ is the filtered one, and $w$ are the weights of the specific normalization. We divide by the sum of the weights to prevent artificial explosion or suppression of the signal's amplitude.

We test two distinct approaches. 
The first is an inverse-variance transform, where each $uv$ cell was divided by its variance $w=\frac{1}{2\sigma_{uv}^2}$\footnote{where we used a factor of 2 to account for the total variance (real and imaginary parts)} (see Sect. \ref{sect:21cm_obs}), resulting in high signal-to-noise cells contributing more to the Fourier transform. 
The second is Wiener filtering where the power spectrum of the final field matches the power spectrum of the intrinsic signal, $w = \sqrt{\frac{1}{2}\frac{P_\text{signal}}{P_\text{signal}+P_\text{noise}}}$, where $P_\text{noise} = 2\sigma_{uv}^2$. This transformation is prior-dependent, as it requires assuming a power spectrum for the intrinsic signal $P_\text{signal}$, that is not observationally known. For this reason, we calculated the median amplitude of the intrinsic 21cm signal in each $uv$ cell, over the prior range of astrophysical parameters in \ct{}, which can be seen in Fig. \ref{fig:appendx:Wiener_fixed}. 
The resulting signal in real space can be seen in the lower-right panel of Fig. \ref{fig:appndx:filtering}, showing more details than the inverse-variance transform but less prominent structure compared to the intrinsic one. 
\texttt{TILING} was indifferent to the choice of these weightings. We chose the inverse variance as a physically motivated, prior-independent choice.

We also present how different choices of thermal noise affect the mock observed SKA signal. 
In the main text, we established a fiducial value of $\sim 1000\text{h}$, as illustrated in the bottom-left panel of Fig. \ref{fig:appndx:noise_levels}.
Figure \ref{fig:appndx:noise_levels} visually demonstrates the differences in the signal in the plane of the sky for $z=8$ averaged over $7.5\,\rm cMpc$ in the LoS direction. The top panel displays two perturbations around the fiducial value (100 and 10000h). The 100h observation shows a significant reduction in visible structure, while the 1000 and 10000 cases are visibly similar. Furthermore, adopting the SKA AA4 configuration does not lead to a visible improvement in morphological clarity, as seen in the bottom right panel.

\section{Network architecture}\label{appdx:NN_architecture}

\begin{table}[htb!]
    \setlength{\tabcolsep}{3pt}
    \centering
    \caption{U-net's architecture based on \citealp{Savchenko+2025}.}
    \label{tab:cnn arch}
    \begin{tabular}{c c c c c}
    \hline\hline
    \noalign{\smallskip}
    
    Layer & Input & I/O dim & Param \# \\ 
    \hline
    \noalign{\smallskip}
    Input data & - & $2 \times 200^3$ &  - \\
    C1           & Input data & $32 \times 200^3 $ & $1,760$  \\
    A1         & previous & "  & - \\
    C2           & " & "  & 27,680 \\
    B1           & " & "  & 64  \\
    A2         & "  & " & -  \\
    \hline
    \noalign{\smallskip}
    Downsample \\
    \hline
    \noalign{\smallskip}
    D1         & "  & $32\times 100^3$ & 8,224 \\
    B2         & "  & " & 64  \\
    A3         & "  & " & "  \\
\hline
\noalign{\smallskip}
    C3         & "  & " & 27,680  \\
    B3         & "  & " & 64  \\
    A4         & "  & " & "  \\
    C4         & "  & " & 27,680  \\
    B4         & "  & " & 64  \\
    A5         & "  & " & "  \\
    \hline
    \noalign{\smallskip}
    Downsample \\
    \hline
    \noalign{\smallskip}
    D2         & "  & $32\times 50^3$ & 8,224  \\
    B5         & "  & " & 64  \\
    A6         & "  & " & "  \\
    \hline
    \noalign{\smallskip}
    C5         & "  & " & 27,680  \\
    B6         & "  & " & 64  \\
    A7         & "  & " & "  \\
    C6         & "  & " & 27,680  \\
    B7         & "  & " & 64  \\
    A8         & "  & " & "  \\

    \hline
    \noalign{\smallskip}
    Upsample \\
    \hline
    \noalign{\smallskip}
    U1         & "  & $32\times 100^3$ & 8,224  \\
    B8         & "  & " & 64  \\
    A9         & "  & " & "  \\
    \hline
    \noalign{\smallskip}
    C7         & concat A2  & $64/32 \times 100^3$ & 55,328  \\
    B9         & "  & " & 64  \\
    A10         & "  & " & "  \\
    C8         & "  & " & 27,680 \\
    B10         & "  & " & 64  \\
    A11         & "  & " & "  \\
    \hline
    \noalign{\smallskip}
    Upsample \\
    \hline
    \noalign{\smallskip}
    U2         & "  & $32\times 200^3$ & 8,224  \\
    B11         & "  & " & 64  \\
    A12         & "  & " & "  \\
    \hline
    \noalign{\smallskip}
    C9         & concat A5  & $64/32 \times 100^3$ & 110,656  \\
    A13         & "  & " & "  \\
    C10         & "  & " & 1,729  \\
    
    \hline

    \noalign{\smallskip}
    \end{tabular}

    \setlength{\tabcolsep}{11pt}
    \begin{tabular}{c c}
    \hline\hline
    \noalign{\smallskip}
    
    Layer & Meaning   \\ 
    \hline
    \noalign{\smallskip}
    C&  Convolution with kernel size 3, stride 1          \\
    A&  LeakyReLU activation function          \\    
    B&  Batch normalization (Group for diffusion)     \\   
    D&   Convolution with kernel size 2, stride 2\\  
    U&   Transpose with kernel size 2, stride 2\\  
    \hline
    \noalign{\smallskip}
    \end{tabular}

    \tablefoot{
    Layers are numbered sequentially. For diffusion, a Fourier time embedding is added before every activation.}
\end{table}

\subsection{U-net}
As explained in the main text, we employ a two-step hybrid model designed to accurately model both the likelihood and the mapping of the high-dimensional space. 
Both steps utilize the same underlying three-level U-net architecture with two skip connections, the details of which are summarized in Table~\ref{tab:cnn arch}. 
From left to right, the table columns show the layer type and the input to each layer. The input is always the output of the preceding layer, except at the skip connections where previous intermediate outputs are concatenated (abbreviated as "concat. A2" and "concat. A5").

For the second step of the hybrid model, this U-net is augmented with a Fourier time embedding added before every activation function. 
We experimented with various placements of this embedding, finding that this configuration yielded the best performance. 
We also explored several architecture modifications: replacing the standard convolutional blocks with residual blocks (following \citealp{Song+2020}) or increasing the number of baseline feature maps from 32 to 64 did not produce noticeable improvements. Consequently, we selected 32 feature maps as the most computationally efficient choice. 
The model was trained using a batch size of 8, distributed across the available GPUs.

\subsection{Hybrid versus single architecture and additional observational scenarios}
\begin{figure}[htb!]
    \centering    \includegraphics[width=0.97\columnwidth]{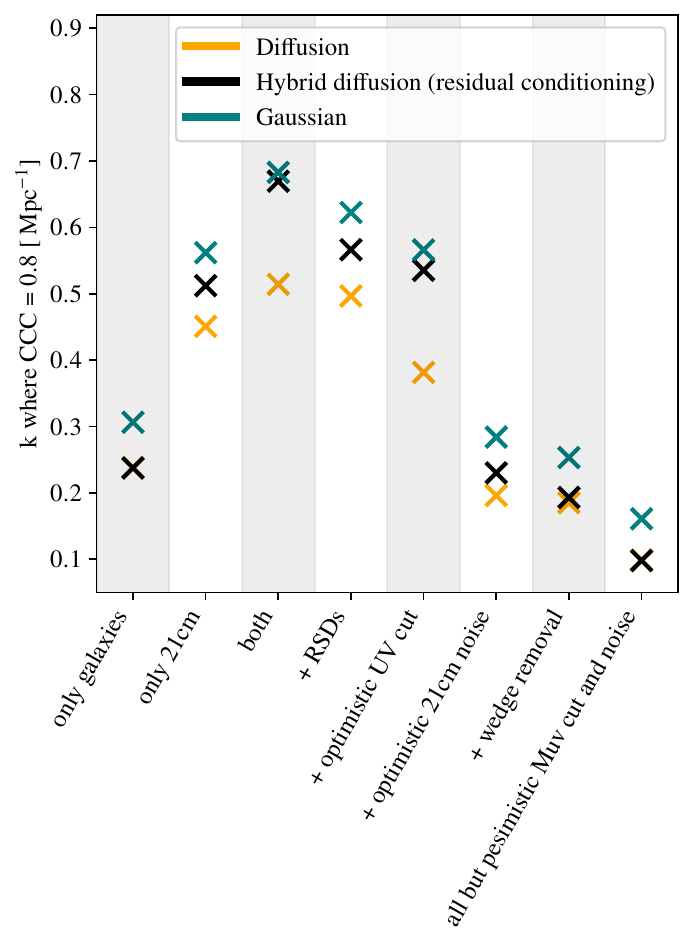}
    \hfill
    \caption{
    Comparison of inferred initial conditions (ICs) across various observational scenarios and inference schemes. The panels display the resulting $\rm k_{08}$ for eight distinct observational scenarios labeled along the x-axis:(i) Galaxy-only single-tracer inference ($M_{\rm h} > 10^9\, M_{\odot}$).(ii) 21cm-only single-tracer inference.(iii) Multi-tracer inference combining both galaxies and the 21cm signal.(iv) Multi-tracer inference including redshift-space distortions (RSDs).(v) Multi-tracer with RSDs and a galaxy luminosity cut of $M_{\rm UV,cut} = -18$.(vi) Same as (v), but with the 21cm signal degraded by the addition of experimental noise corresponding to a $\sim$64000h SKA AA* observation.(vii) Same as (vi), after removing the 21cm foreground wedge.(viii) Same as (vii), under pessimistic constraints ($M_{\rm UV,cut} = -19$ and a $\sim$6400h observation time).Colors indicate three different inference schemes, with \texttt{TILING} shown in black. Across all scenarios, \texttt{TILING} either outperforms or matches simple diffusion (orange). Teal markers represent a model trained only on the first stage of the hybrid framework; while it displays an elevated $\rm k_{08}$, this does not capture the full performance, as other key statistics—including the power spectra, bispectra, and PDF (yielding a skewness and kurtosis of $-0.15$ and $-0.13$, respectively)—deviate significantly from the target ICs. Furthermore, \texttt{TILING} outperforms the NGPE in CCC when evaluating its MAP assignment. The first four columns emphasize the substantial complementary information embedded in the combined galaxy and 21cm fields, which becomes inaccessible under the realistic observational and noise limitations shown in the remaining panels}
    \label{fig: CCC comparison}
\end{figure}

Here we show the resulting ICs inference using each part of our two-step hybrid and demonstrate its increased performance. In Fig. \ref{fig: CCC comparison} the resulting $\rm k_{08}$ is shown from eight observational scenarios labeled across the x axis. From left to right the observational scenarios are: (i) a single tracer inference only considering galaxy observations considering all galaxies residing in halos of $M_{\rm h}>10^9\, M_{\odot}$; (ii) a single-tracer scenario considering only the intrinsic 21cm signal; (iii) a multi-tracer scenario considering both of the previously mentioned tracers; (iv) the same as the previous scenario adding the effect of RSDs; (v) the same as before but with an $M_{\rm UV,cut}=-18$; (vi) same as before with the 21cm signal being degraded due to noise addition corresponding to $\sim$64000h SKA AA* observation; (vii) same as before after removing a foreground wedge from the 21cm signal and (viii) same as before but with a worse $M_{\rm UV,cut}$ of -19 and a $\sim$6400h observation.
By looking at the first four columns it is evident that there is significant complementary information in the galaxy and 21cm fields, which is inaccessible under the realistic observational scenarios detailed in the main text under current observational limitations.

Three different inference schemes are shown with different colors, with {\tt TILING} shown in black.
In all cases \texttt{TILING} outperforms simple diffusion (orange) or is equivalent.
The teal-colored markers correspond to a model trained only with the first part of the model showing a better $\rm k_{08}$. However, this performance does not capture the full picture as other statistics such as the power spectra, the PDF (with skewness and kurtosis of -0.15 -0.13) and bispectra were deviating significantly from the target ICs. 
Furthermore, \texttt{TILING} outperforms the NGPE in the CCC when its MAP is considered (see also \citealp{Savchenko+2025} for the MAP behavior compared to the samples on the CCCs).

\section{Statistical tests of recovery}\label{appndx:sec:statistical_tests}
In Sect. \ref{sec:evolved_fields} we performed recovery tests in the evolved space at $z=8$. 
Here, we expand upon those results by conducting rigorous statistical diagnostics. These include field-level Gaussianity checks and coverage tests applied directly to the reconstructed ICs, alongside validation of the HMFs derived from re-simulating the posterior samples of the fiducial model.

\subsection{Gaussianity and PNGs}\label{sec:appndx:tests_gauss}
Throughout this work, we assume purely Gaussian ICs. Hence, our reconstruction should be able to preserve this Gaussianity, even though nowhere in the loss function or validation it was explicitly stated.
 
\begin{figure}[htb!]
    \centering
    \includegraphics[width=0.96\columnwidth]{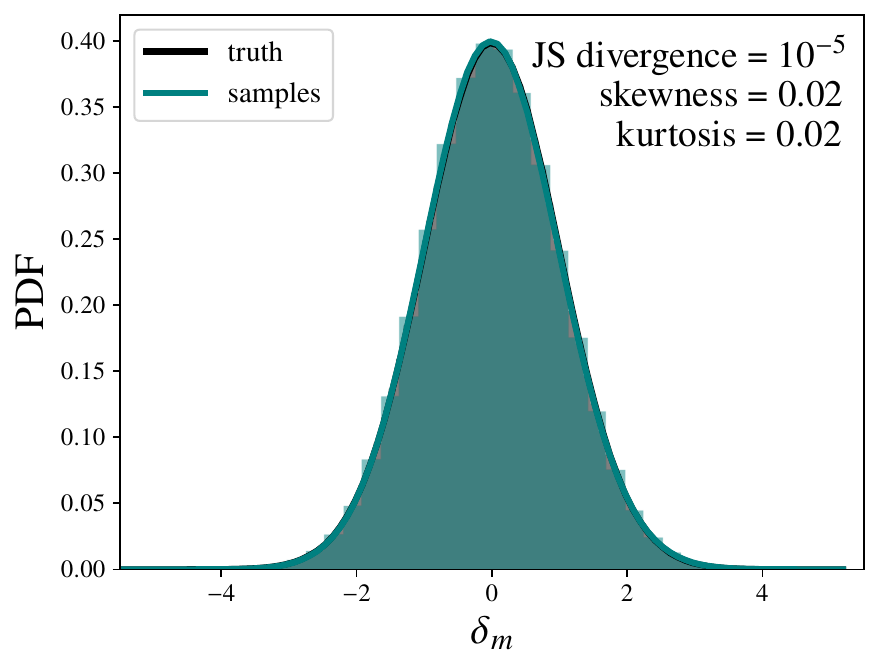}
    \hfill
    \caption{Histogram of all samples (teal) versus the test set for the true distribution (black). The base-two Jensen–Shannon (JS) divergence and the skewness and kurtosis are reported. Standard deviations of their calculations are ignored as we do not do bootstrapping. Given the minimal value in all metrics and especially the JS divergence which is naturally bounded in [0,1] we conclude the distributions are statistically indistinguishable.}
    \label{fig:appndx:PDF}
\end{figure}

 We evaluate the Gaussianity requirement in Fig. \ref{fig:appndx:PDF} by plotting the voxel PDF sampled from a test-set  sample. Apart from the distributions being visually closely aligned, we use the three metrics on the panel to quantify their similarity.
First, the skewness and kurtosis (the third and fourth standardized moments) are negligible ($=0.02$ for both), confirming the samples are Gaussian for all practical purposes. Second, we further quantify this minimal deviation using the base-two Jensen-Shannon divergence (JS; e.g., \citealp{linDivergenceMeasuresBased1991, Dorent+2025}), which offers a robust framework to quantify this minimal deviation. For two probability distributions, $P(x)$ and $Q(x)$, the JS is defined as:
\begin{equation}
    \text{JS}(P \parallel Q) = \frac{1}{2} \left[ D_{\text{KL}}(P \parallel M) + D_{\text{KL}}(Q \parallel M) \right],
    \end{equation}
    where $M(x) = \frac{1}{2}[P(x) + Q(x)]$ represents the average distribution, and the (base-two) Kullback-Leibler divergence $D_{\text{KL}}$ is given by:
    \begin{equation}
    D_{\text{KL}}(P \parallel M) = \sum_x P(x) \log_2 \frac{P(x)}{M(x)}.
\end{equation}
The metric is bounded within $[0,1]$, where a value of 0 indicates identical distributions and 1 signifies completely disjoint profiles. Notably, our reconstructed fields yield a JS of $\sim 10^{-5}$, indicating that the recovered PDF is practically identical to the true distribution.

Beyond validating the pipeline's fidelity, the precision with which the framework preserves this baseline Gaussianity serves as a critical indicator of its capacity to detect PNGs. As noted in the introduction, field-level IC reconstruction offers a powerful avenue for placing constraints on early-universe non-Gaussian physics.  
To evaluate the framework's capability to constrain PNGs, we measure selected three-point statistics of the fields.

The standard bispectrum, $B(\text{k}_1, \text{k}_2, \text{k}_3)$, measures the three-point correlation function in Fourier space. For a discrete field, it is estimated by averaging over all valid triangular configurations that satisfy the closed-triangle condition $\boldsymbol{k}_1 + \boldsymbol{k}_2 + \boldsymbol{k}_3 = \mathbf{0}$:
\begin{equation}
    B(\text{k}_1, \text{k}_2, \text{k}_3) = \frac{1}{N_{\text{tri}}} \sum_{\boldsymbol{k}_1, \boldsymbol{k}_2, \boldsymbol{k}_3} \delta(\boldsymbol{k}_1) \delta(\boldsymbol{k}_2) \delta(\boldsymbol{k}_3),
\end{equation}
where $N_{\text{tri}}$ is the number of fundamental triangles falling within the specified $\rm k$-shells.
Hereafter, we refer to the bispectrum as the reduced bispectrum, which has a weaker dependence on cosmology and scale, and is defined as:
\begin{equation}
    Q(\text{k}_1, \text{k}_2, \text{k}_3) = \frac{B(\text{k}_1, \text{k}_2, \text{k}_3)}{P(\text{k}_1)P(\text{k}_2) + P(\text{k}_2)P(\text{k}_3) + P(\text{k}_1)P(\text{k}_3)},
\end{equation}
where $P(\rm k)$ represents the matter power spectrum at wavenumber $\rm k$.

\begin{figure}[htb!]
    \centering
    \includegraphics[width=0.97\columnwidth]{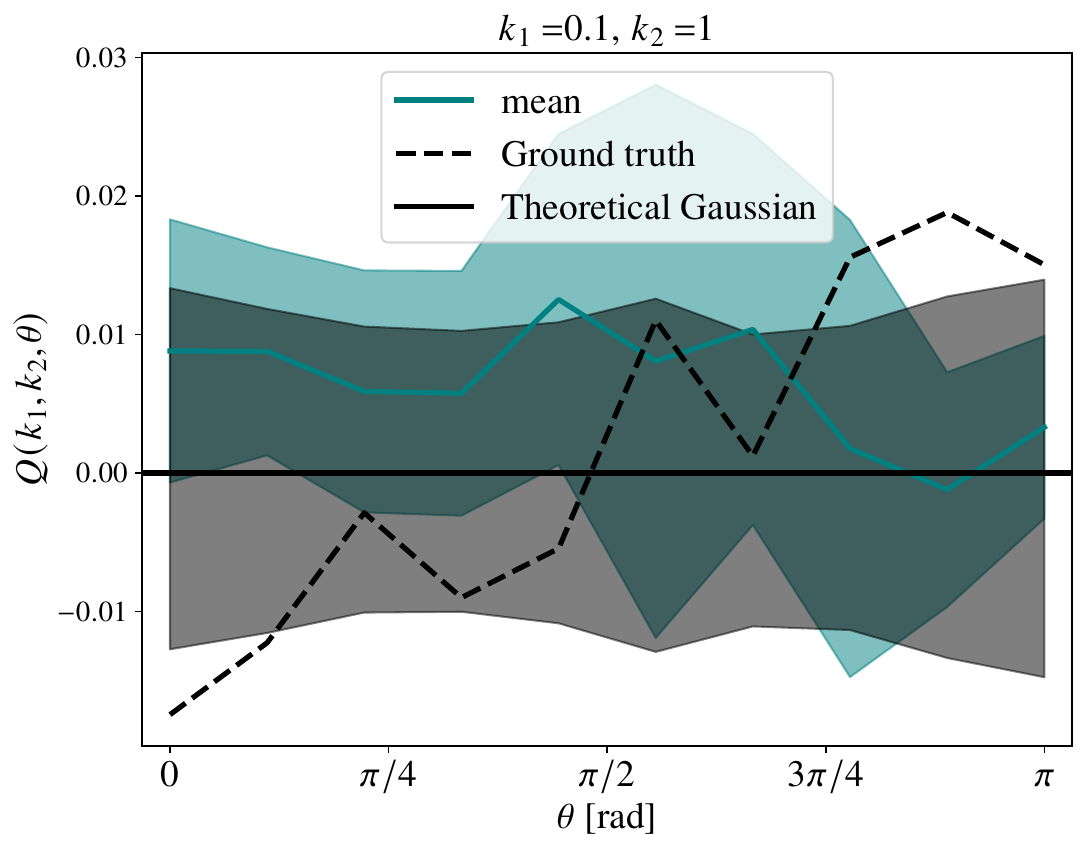}
    \includegraphics[width=0.97\columnwidth]{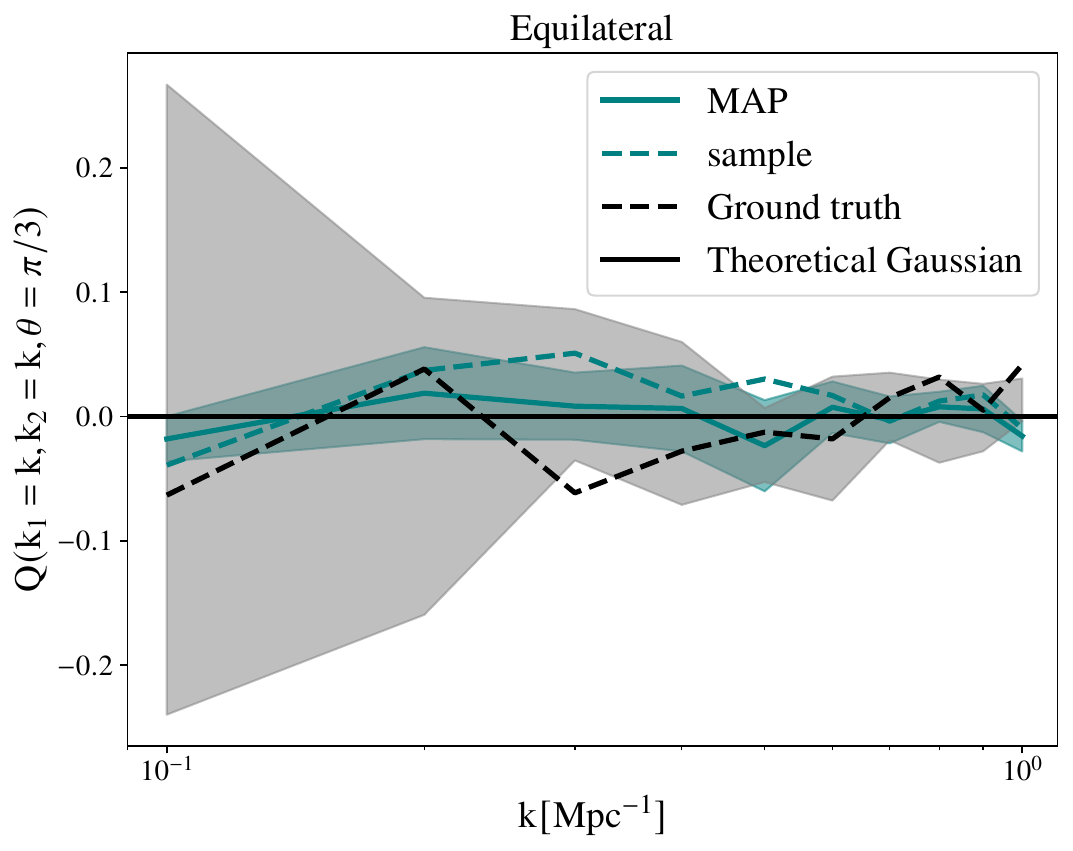}
    \hfill
    \caption{Reduced bispectra of 10 ICs samples on the linearly extrapolated overdensity field, given one test-set realization. The black shaded regions indicate the empirical standard deviation (around 0) of the bispectra under Gaussianity, calculated from the whole 2000 simulations suite. With the black dashed line, the exact test-set realization bispectrum is shown. The samples' mean and $1\sigma$ is shown as the teal-colored line and shaded regions respectively. In both triangle configurations, the mean of the samples is located inside the 1$\sigma$ region under Gaussianity. 
    }
    \label{fig:appndx:bispectra}
\end{figure}

Using the \texttt{Pylians3} package \citep{Navarro+2018}\footnote{\url{https://github.com/franciscovillaescusa/Pylians3}}, we compute bispectra for a random test-set realization along with 10 posterior samples displayed in Fig.~\ref{fig:appndx:bispectra}. 
Specifically, we evaluated two distinct triangular configurations: the equilateral bispectrum ($\rm k_1 = k_2 = k_3 = k$) as a function of wavenumber $\rm k$ (upper panel), and the angular dependence of the bispectrum for a fixed pair of large- and small-scale modes (lower panel). For the latter, fixing the magnitudes of $\rm k_1$ and $\rm k_2$ along with their angular separation $\theta$ uniquely determines the third vector's magnitude, $\rm k_3$, via the closed-triangle condition. 
Because of the simulation box's discreteness and cosmic variance, the bispectrum exhibits inherent scatter even for purely Gaussian fields. To establish this baseline, we calculate the empirical mean and variance of the Gaussian bispectra across the entire 2000 realizations of the simulation suite. These are represented by the gray shaded regions in Fig.~\ref{fig:appndx:bispectra}, which define the expected scatter boundaries under Gaussianity.

The true bispectrum realization is shown as the black dashed line while the samples' mean and 1$\sigma$ is shown with the teal solid line and shaded regions respectively. For both configurations, the posterior samples demonstrate statistical consistency as their empirical mean remains well within the $1\sigma$ envelope of the expected Gaussian scatter.\footnote{
We note that this is a potential indication of PNG recovery but it remains a sanity check in our idealized Gaussian scenario.}

\subsection{Halo mass function}

\begin{figure}[htb!]
    \centering
    \includegraphics[width=0.99\linewidth]{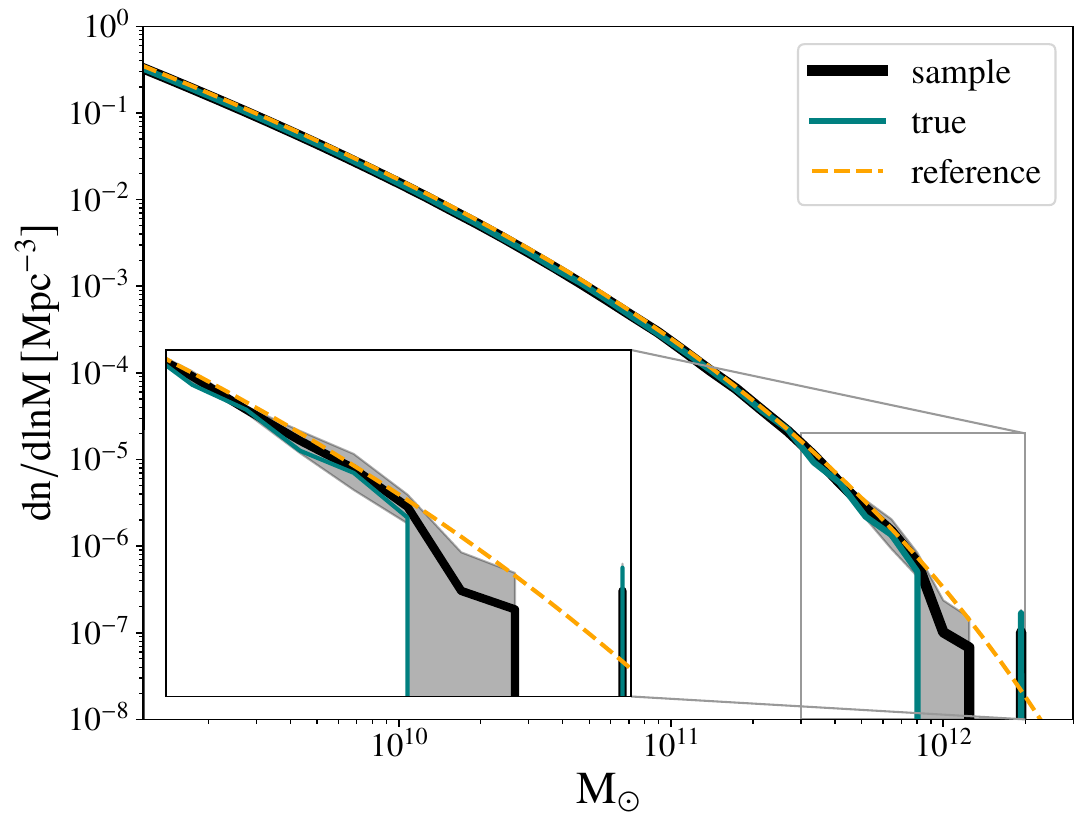}
    \caption{Halo mass function (HMF) comparison at $z=8$. The true HMF for a single test-set realization is shown in teal, plotted alongside the analytical Sheth-Tormen HMF reference (dashed line). Solid black lines and surrounding shaded regions represent the mean and $1\sigma$ variance across 10 posterior samples, respectively. The teal colored peak in the high-mass end indicates a singular, massive halo. The posterior samples closely trace both the true distribution and the reference model at lower mass scales, with variance increasing toward the higher-mass regime but still being within $1\sigma$ and successfully reconstructing the massive halo.}
    \label{appndx:fig:hmf}
\end{figure}

In Sect. \ref{sec:evolved_fields} we showed the reconstructed density field and 21cm field at $z=8$.  In Fig. \ref{appndx:fig:hmf}, we provide the halo mass function (HMF) as an additional summary of re-simulated quantities. The true HMF of one test-set realization (teal) is displayed along with the mean and $1\sigma$ of 10 samples (solid lines and shaded regions respectively) and a Sheth-Tormen HMF (\citealp{Sheth+1999}) as a reference.
The samples closely match the truth and the reference for the small masses, as expected, due to the correct statistics of the ICs explored in the previous paragraphs.
At higher masses, the samples showcase more uncertainty, associated with the densest regions and their dynamical histories.
Notably, the single peak in the true HMF in the high-mass end corresponds to one halo under the simulation setup. This halo is successfully reconstructed in most of the posterior samples as indicated by the black line.

\subsection{Posterior coverage tests}\label{appndx:sec:coverage}

To further verify the validity of our posteriors across different test-set realizations (in addition to the PSRs and CCCs detailed in Sect. \ref{sec:dependence_on_properties}), we perform coverage tests.

Coverage tests measure how well an approximate posterior $\hat{p}(\theta|d)$ (being a product of an implicit likelihood formalism such as in our case) traces the exact $p(\theta|d)$ under a single idea:  how well the posterior traces the prior when data are drawn from the prior predictive distribution. 
That is, the prior- and data-averaged posterior should equal the prior: 
\begin{equation}\label{eq:appndx:coverage}
    p(\theta) = \int\int  \hat{p}(\theta|\tilde{d}) p(\tilde{d}|\tilde\theta)p(\tilde\theta)\text{d}\tilde\theta\text{d}\tilde{d}.
\end{equation}
Equation \ref{eq:appndx:coverage} is an identity when the approximate posterior $\hat{p}(\theta|d)$ is replaced with the exact posterior $p(\theta|d)$.
To evaluate self-consistency under this identity, coverage tests check whether the ground-truth parameters $\tilde{\theta}$ fall \textit{within} a credible region of a given $a\%$ credibility level of the approximate posterior exactly $a\%$ of the time (yielding an empirical cumulative distribution
function; ECDF). 
In other words, they check whether the ground-truth parameters $\tilde{\theta}$ fall \textit{at the edge} of an $a\%$ credible region with a uniform probability across all credibility levels (hereafter referred to as ``coverage per bin'' or EPDF, serving as the empirical probability density function counterpart of the ECDF).

The definition of a credible region is non-unique across implementations, as it relies on a credible region generator $\mathcal{G}(\hat{p}, \alpha) = U$ (where $U$ is a subset of the parameter domain) whose sole defining condition is:
\begin{equation}
    \int_{U} \hat{p}(\theta|d)\,\mathrm{d}\theta = \alpha.
\end{equation}
Simulation-based calibration (SBC; \citealp{Talts+18}) is essentially using a ``scanning'' credible region generator (i.e., the ranks) for  marginal distributions (and so limited to 1D posteriors), whereas in tests
of accuracy with random points (TARP; \citealp{Lemos+2023}) they prove that a positionable credible region generator (depending on d) provides a necessary and sufficient condition for good coverage in arbitrary dimensions, but under implementation constraints while affected by the curse of dimensionality.

\begin{figure}[htb!]
    \centering \includegraphics[width=0.90\columnwidth]{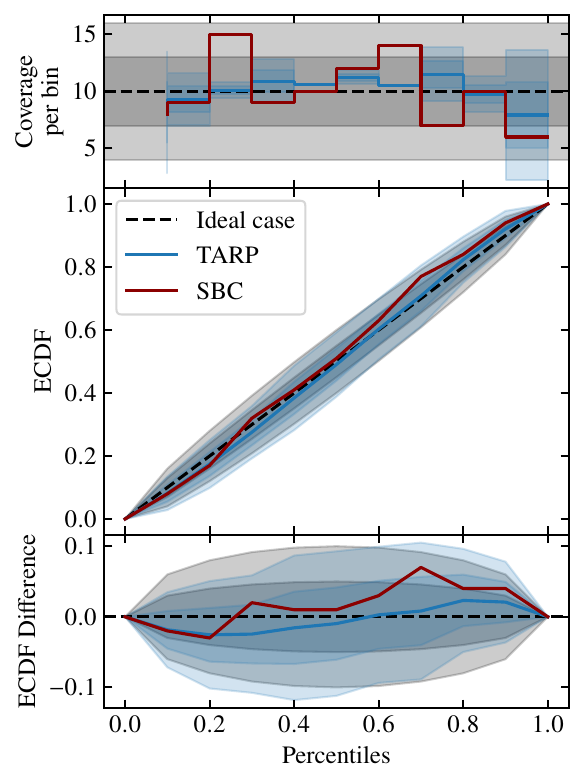}
    \hfill
    \caption{
    Coverage diagnostics evaluating posterior calibration using SBC (red) and TARP (blue) across 100 test-set simulations ($1000$ samples per effective ground truth). Top panel: EPDF. Middle panel: ECDF. Bottom panel: ECDF residual relative to the ideal $1:1$ reference line. All metrics are evaluated across $b = 10$ equiprobable bins. Gray shaded bands indicate the theoretical $1\sigma$ and $2\sigma$ sampling variance expected from a Binomial noise distribution. For TARP, blue shaded regions denote the $1\sigma$ and $2\sigma$ error bounds estimated via bootstrapping with replacement across the 100 realizations. Both SBC and TARP stay well within the expected $2\sigma$ bounds, confirming that our fiducial inferred posterior is well-calibrated.}
    \label{fig: coverage_tests}
\end{figure}

However in our case of $200^3\sim 10^{7}$ parameter reconstruction it is not possible to examine each individual parameter. 
Furthermore, for a coverage test multiple samples from multiple realizations are required. Because of the dimensionality of the problem and expensive sampling,
we make the following approximation: voxels in the realizations are different parameters during inference, but here we assume they can be treated as one single effective parameter $\theta_{\rm eff}$ due to the ergodic nature of the ICs\footnote{This approximation holds as long as our effective samples are not correlated.}. To this end, we sample once for 100 test-set realizations and randomly select $100$ voxel locations to act as different effective parameter ground truths (one per test-set realization). In each test-set realization, we find the first 1000 values that are closest to the corresponding $\theta_{\rm eff}$.
Then, we treat the values of the sample in these voxel positions as the effective samples of $\theta_{\rm eff}$. 
This results in 100 ground truths with 1000 samples in each case, with which we perform coverage tests.

In Fig. \ref{fig: coverage_tests} both SBC and TARP are shown (red and blue respectively) as the EPDF (upper panel), ECDF (middle panel) and the ECDF difference from the ideal 1:1 line (lower panel). In all cases, the results are distributed into $b = 10$ bins. 
In all panels, the expected errors due to sampling variance are shown as the gray shaded regions (1 and 2$\sigma$), and are calculated based on the uniformity condition of the EPDF. That is, for $n=100$ simulations, for each bin, this is a series of Bernoulli trials,  i.e. it follows a binomial distribution, where $E(X) = np$ and $Var(X)=np(1-p)$ where $p=1/b$ and $X$ the number of samples per bin. The same holds for the ECDF, just by changing the probability of success to $p=i/b$ where $i$ is the $i$th bin. 
For TARP, we also show the estimated 1 and 2$\sigma$ regions from bootstrapping (blue shaded regions), that is, performing the same TARP test 100 times, for 100 selected ground truths with resampling.
Both the SBC and TARP fall well within the $2\sigma$ region, indicating a well calibrated posterior.

\end{appendix}

%%%%%%%%%%%%%%%%%%%%%%%%%%%%%%%%%%%%%%%%%%%%%%%%%%

\end{document}